\documentclass[arxiv, preprint]{imsart}

\RequirePackage[OT1]{fontenc}
\RequirePackage{amsthm,amsmath,amssymb}
\RequirePackage{natbib}
\RequirePackage{enumerate, graphicx, verbatim, multirow, color}

\startlocaldefs
\numberwithin{equation}{section}
\theoremstyle{plain}
\newtheorem{theorem}{Theorem}[section]

\newtheorem{lemma}[theorem]{Lemma}
\newtheorem{remark}[theorem]{Remark}
\newtheorem{assumption}[theorem]{Assumption}
\newtheorem{condition}{Condition}
\endlocaldefs

\newcommand{\by}{\mathbf{y}}
\newcommand{\bY}{\mathbf{Y}}
\newcommand{\bI}{\mathbf{I}}

\def\pr{\textsf{P}} 
\def\ep{\textsf{E}} 
\def\Cov{\textsf{Cov}} 
\def\Var{\textsf{Var}} 

\graphicspath{{../code/} {../code/2018/} {../authorship/} {../networkData/}}

\begin{document}

\begin{frontmatter}
\title{Sequential change-point detection based on nearest neighbors} 
\runtitle{Sequential detection by nearest neighbors}

\begin{aug}
\author{\fnms{Hao} \snm{Chen} \ead[label=e1]{hxchen@ucdavis.edu}}

\runauthor{Hao Chen}

\affiliation{University of California, Davis}

\address{Department of Statistics\\
University of Calfornia, Davis\\
One Shields Avenue \\
Davis, Calfornia 95616 \\
USA \\
\printead{e1}\\
\phantom{E-mail:\ }}

\end{aug}

\begin{abstract}
We propose a new framework for the detection of change-points in online, sequential data analysis.  The approach utilizes nearest neighbor information and can be applied to sequences of multivariate observations or non-Euclidean data objects, such as network data.  Different stopping rules are explored, and one specific rule is recommended due to its desirable properties.  An accurate analytic approximation of the average run length is derived for the recommended rule, making it an easy off-the-shelf approach for real multivariate/object sequential data monitoring applications.  Simulations reveal that the new approach has better performance than likelihood-based approaches for high dimensional data.  The new approach is illustrated through a real dataset in detecting global structural changes in social networks.

\end{abstract}



\end{frontmatter}

\section{Introduction}

Sequential change-point models are widely used in many fields to detect events of interest as data are generated.  One of its early applications is in quality control where a summary statistic reflecting a manufacture process is monitored over time.  When the statistic begins to exhibit values that are unlikely to be achieved by random fluctuations, there is a high probability that something went wrong and an investigation is needed.  Therefore, it is important to detect the change-point, the time when the event of interest happens, as soon as possible if it occurs, while keeping the false discovery rate low; refer to monographs \cite{wald1973sequential}, \cite{siegmund1985sequential} and \cite{tartakovsky2014sequential} for more background information.  

Sequential change-point detection has been extensively studied for univariate data, that is, for data where the observations are scalar at each time point.  However, many recent applications involve the detection of change-points over a sequence of multivariate, or even non-Euclidean, observations.  Following are some motivating examples.


\begin{description}
\item[Multiple sensor framework:] In a sensor network, hundreds or thousands of sensors are deployed to detect events of interest.  For example, hundreds of monitors are placed worldwide to detect solar flares, which are large energy releases by the Sun that can affect Earth's ionosphere and disrupt long-range radio communication \citep{kappenman2012perfect, qu2005automatic}.  Often, the structure of the sensor network can be used to boost the power of the detection.  Then, each observation can be viewed as a vector with some structures among its elements that reflect the spacial information of the sensors.
\item[Social network evolution:]  Technological advances provide us with rich resources of social network data, such as networks constructed by Facebook friendship relations, email communications, phone calls, or online chat records.  The detection of abrupt events, such as shifts in network connectivity, dissociation of communities, or formation of new communities, can be formulated as a change-point problem.  Here, the observation at each time point is a graphical encoding of the network. 
\item[Epidemic disease outbreak:] It is important to detect the emergence of new infectious diseases as early as possible to prevent their spreadings.  In United States, the current practice is that the Center for Disease Control gathers data from hospitals and then integrates information together to tell if there is an outbreak.  This process usually takes weeks to draw conclusions.  Researchers have tried to incorporate other information, such as online searches on disease related topics and climate information, which had success in shortening the prediction lag time for flu outbreaks \citep{yang2015inference, yang2015accurate}.  It can be foreseen that, in the future, information from multiple sources will be used to predict disease outbreaks.  Then, each observation can be quite complicated and may include hospital admission rates, online search frequencies on related topics, personal posts on related symptoms, and whether information. 

\item[Image analysis:]  Image data are collected over time in many areas.  It is of tremendous interest to automatically detect abrupt events, such as security breaches from surveillance videos or extreme weather conditions, e.g., storms, from climatology.  In these applications, the data at each time point is the digital encoding of an image.  

\end{description}

In all of these examples, the problem can be formulated in the following way: We denote the data sequence by $\{\bY_i\}, i=1,2,\dots,n,\dots$, indexed by time or some other meaningful orderings.  Here, $\bY_i$'s can be vectors, networks, or images.  The sequence is identically distributed as $F_0$ until a time $\tau$ the distribution changes abruptly to $F_1$:
$$\bY_i\sim F_0, \quad i=1, \dots, \tau-1,$$
$$\bY_i\sim F_1, \quad i=\tau, \tau+1,\dots,$$
where $F_0$ and $F_1$ are two different probability measures.

There is a burst of works recently on the change-point detection in multiple sequences where the sequences are assumed to be independent, such as in multiple sensor framework where the sensors are assumed to be indenpendent.  These works also in general assume the observations over time are independent.  
Some nice algorithms and theorems have been developed under these assumptions.  For example,  \cite{tartakovsky2008asymptotically} and \cite{mei2010efficient} studied statistics that sum signals over all streams with further assumptions that the density functions before and after the change are known and the change happens to all streams at the same time.  \cite{xie2013sequential} and \cite{chanwalther2015optimal} allow the change only happen to a subset of the data streams under the assumption that $F_0$ and $F_1$ are multivariate normal distributions with identity covariance.  The latter paper also studied the optimality of several statistics.  These statistics are useful if the assumptions under which the statistic was developed hold for the data.  However, in many applications, it would be too stringent to assume that all data streams are indenpendent.

To another end, the change-point detection problem for dynamic network data is gaining more and more attentions.  A number of works have been done if the networks are generated in some specific ways.  For example, \cite{heard2010bayesian} developed Bayesian methods by modeling each pair of nodes independently and they modeled the communications between nodes over time as a counting process with the increments of the process following a Bayesian probability model.  A multinomial extension that relaxed the independence assumption among pairs of nodes was also studied by the authors.  \cite{wang2014locality} considered the setting that the series of networks are generated by a stochastic block model with the block membership of the vertices fixed across time.  They use locality-based scan statistic to find change-point where the connectivity probability matrix varies.  Again, these methods are useful if the data do satisfy the assumptions, while these assumptions could be too specific for many applications.




In this paper, we describe a nonparametric framework to approach the problem.  This framework can be applied to data in arbitrary dimension and to non-Euclidean data, with a general, analytic formula for false discovery control.  The proposed method adopts the idea of making use of similarity graphs, such as nearest neighbors, among the observations in \cite{chen2015graph}.  

In the following, we do not impose specific assumptions on $F_0$ or $F_1$.  However, we assume the observations over time are independent.  When there is weak dependence over time, the graph-based approach could still provide meaningful results for change-point analysis as shown in \cite{chen2015graph}.  Also, the independence assumption is a natural starting point for more sophisticated models that consider dependency over time.

\cite{chen2015graph} studied the problem of \emph{offline} change-point detection, where all observations are completely observed at the time when data analysis is conducted.  However, in many applications, it is desirable to detect change-points on the fly.  There are both theoretical and computational challenges to extend the method in \cite{chen2015graph} to the \emph{online} framework.  In particular, adding new observations usually changes the similarity structure among existing observations, when the most similar observation for an existing observation may be changed to the newest observation.  This makes the theoretical analysis on false discovery control much harder as it requires an analysis of the dynamics of similarity structural change when new observations are added.  

In this paper, we consider the similarity structure represented by nearest neighbors (NN). 
We studied the dynamics in NN updates as new observations are added.  It turns out that the characterization of a small number of events, in particular, the updates of mutual NNs and shared NNs, and all three-way interactions among the NN relations, could capture the majority of the dynamics (see Section \ref{sec:EDD} for details).  This makes the task tractable.  We can also easily implement the method for real data applications.


The rest of the paper is organized as follows:  In Section \ref{sec:twosampletest}, we briefly review a two-sample test based on NNs, which is a building block for the change-point analysis.  In Section \ref{sec:cp}, we discuss details of the proposed detection method and three stopping rules.  We recommend the use of the stopping rule that relies on recent observations for its desirable properties.  In Section \ref{sec:ARL}, we study the updating dynamics of NNs and derive an analytic formula for false discovery control that is accurate for finite samples.  In Section \ref{sec:EDD}, we compare the proposed method to parametric methods for multivariate data.  We illustrate the proposed method on a real dataset in Section \ref{sec:realdata}.  In Section \ref{sec:discussion}, we briefly discuss the choice of the number of nearest neighbors, the performance of the proposed method on gradual changes, and possible extensions of the method to other similarity graphs. 

\section{A brief review of the two-sample test on $k$-NN}
\label{sec:twosampletest} 
 In this section, we review the two-sample test on $k$-NN proposed by 
 \cite{schilling1986multivariate} and \cite{henze1988multivariate}.  Here, $k$ is a fixed integer. 
 Let $k$-NN be the directed graph with the pooled observations as the nodes and each node points to its first $k$ NNs.  It is assumed that the observations are distinct with uniquely defined neighbors. (This happens with probability 1 if $\bY_i$'s follow continuous multivariate distributions and the Euclidean distance is used.)

Let $\{\bY_1,\dots, \bY_{n_1}\}$ and $\{\bY_{n_1+1},\dots \bY_{n_1+n_2}\}$ be random samples from two populations, and let $n=n_1+n_2$ be the total sample size.  For any event $x$, let $\bI(x)$ be the indicator function that takes value 1 if $x$ is true or 0 if otherwise.  Let $$b_{ij} = \bI((i\leq n_1, j>n_1) \text{ or } (i>n_1, j\leq n_1)),$$  then $b_{ij}$ is the indicator function that  $\bY_i$ and $\bY_j$ belong to different samples.
We want to test whether these two population distributions are the same or not.  Let 
\begin{align*}
A_{ij}^{(r)} & = \bI(\bY_j \text{ is the $r$th nearest neighbor of } \bY_i), \quad A_{ij}^+ =\sum_{r=1}^k A_{ij}^{(r)}.
\end{align*}
Then $A_{ij}^+$ is the indicator function that $\bY_j$ is among the first $k$ NNs of $\bY_i$.  We have $A_{ij}^+ \in \{0,1\}$ and $\sum_{j=1}^n A_{ij}^+ = k, 1\leq i\leq n$.
Then 
$$\sum_{i=1}^n\sum_{j=1}^n A_{ij}^+ \ b_{ij}
\equiv \sum_{i=1}^n\sum_{j=1}^n A_{ji}^+\ b_{ij}
$$
is the number of edges in the $k$-NN that connect between the two samples.  

Expressing in a more symmetric way, we have
\begin{align}
\sum_{i=1}^n\sum_{j=1}^n (A_{ij}^+ + A_{ji}^+)\, b_{ij}
\end{align}
being twice the number of edges in the $k$-NN that connect between the two samples.
Given the observations $\bY_i=\by_i, 1\leq i\leq n$, the test statistic is $$\sum_{i=1}^n\sum_{j=1}^n (a_{ij}^+ + a_{ji}^+)\, b_{ij},$$ where $a_{ij}^+ = \sum_{r=1}^k a_{ij}^{(r)}$ with $a_{ij}^{(r)} = \bI(\by_j \text{ is the $r$th nearest neighbor of } \by_i)$.  In \cite{schilling1986multivariate} and \cite{henze1988multivariate}, the authors proposed to reject the null hypothesis of no difference if the test statistic is significantly \emph{smaller} than its expectation under the permutation null distribution.  The rationale is that, if the two samples are from the same distribution, they are well mixed and are likely to find their nearest neighbors from the other sample.  So if the observations tend to not find nearest neighbors from the other sample, they are from different distributions.

We denote the random variable under the permutation distribution as follows:
Let $B_{ij} = b_{\textbf{P}(i)\textbf{P}(j)}$ be the indicator function that $\bY_i$ and $\bY_j$ belong to different samples under random permutation.  Here, $\textbf{P}(i)$ is the index of $\bY_i$ under permutation.  Let
\begin{align}
X = \sum_{i=1}^n\sum_{j=1}^n (a_{ij}^+ + a_{ji}^+)\, B_{ij}.
\end{align}
Then its expectation and variance are
\allowdisplaybreaks
\begin{align*}
\ep(X) & =  \frac{4k\, n_1\, n_2}{n-1}, \\
\Var(X) &  = \frac{4\, n_1\, n_2}{n-1}\left(h(n_1,n_2)\left(\frac{1}{n}\sum_{i,j=1}^n a_{ij}^+ a_{ji}^++k-\frac{2k^2}{n-1} \right) \right. \\
& \quad \quad \quad \quad \quad \quad \ \left. + (1-h(n_1,n_2))\left(\frac{1}{n}\sum_{i,j,l=1}^n a_{ji}^+ a_{li}^+-k^2\right) \right),
\end{align*}
where 
$ h(n_1,n_2)  = \tfrac{4(n_1-1)(n_2-1)}{(n-2)(n-3)}.$
It has been shown that 
$$\frac{X-\ep(X)}{\sqrt{\Var(X)}}$$
converges to the standard normal distribution under the null hypothesis as long as $n_1/n_2 \rightarrow \lambda \in(0,\infty)$ for multivariate data \citep{schilling1986multivariate, henze1988multivariate}. 


\section{Sequential change-point detection based on $k$-NN}
\label{sec:cp}

We use $$\bY_1,\bY_2,\dots,\bY_n,\dots$$ to denote the data sequence, where $\bY_n$ is the observation at time $n$.  In the following, we assume that we have a well defined norm $\|\cdot\|$ on the sample space such that the distance between two observations $\by_i$ and $\by_j$ can be calculated as $d(\by_i, \by_j) = \|\by_i-\by_j\|$. We also assume that the observations are distinct points in the sample space and have uniquely defined nearest neighbors.  In the following, $k$ is fixed.  The choice of $k$ is briefly discussed in Section \ref{sec:conclusion}.

We assume that there are $N_0$ historical observations with no change-point. That is, $\bY_1,\bY_2,\dots,\bY_{N_0}$ follow the same distribution.  This can be determined from prior information or we can use offline change-point detection methods to test whether there is any change-point among the first $N_0$ observations, such as the method in \cite{chen2015graph}.  We begin our test from observation $N_0+1$.


For any $n$, $1\leq i,j\leq n$, let 
$$ A_{n,ij}^{(r)} = \bI(\bY_j \text{ is the $r$th NN of } \bY_i \text{ among the first $n$ observations}),. $$
Then $A_{n,ij}^+  = \sum_{r=1}^k A_{n,ij}^{(r)}$ is the indicator function that $\bY_j$ is one of the first $k$ NNs of $\bY_i$ among the first $n$ observations.

We can perform a two-sample test for each $t\in\{1,\dots,n-1\}$ with one sample being the observations before $t$ and the other sample being the observations between $t$ and $n$. Define
$$b_{ij}(t,n) = \bI((i\leq t, t<j\leq n) \text{ or } (t<i\leq n, j\leq t)),$$
and $B_{ij}(t,n)=b_{\textbf{P}_n(i) \textbf{P}_n(j)}(t,n)$, where $\textbf{P}_n(\cdot)$ is a random permutation among the first $n$ indices.  Let
$$R(t,n) = \sum_{i=1}^n\sum_{j=1}^n (A_{n,ij}^+ +A_{n,ji}^+) B_{ij}(t,n).$$
We use $\by_i$'s to denote the realizations of $\bY_i$'s, and let
$$Z_{|\by}(t,n) = -\frac{R(t,n)-\ep(R(t,n))}{\sqrt{\Var(R(t,n)|\by)}}.$$
Note that $\ep(R(t,n)|\by) = \ep(R(t,n))$.

If a change-point $\tau>N_0$ occurs in the sequence, we would expect $Z_{|\by}(t,n)$ to be \emph{large} (notice the negative sign in the standardization) when $n>\tau$ and $t$ close to $\tau$.  In the following, we consider three stopping rules:
\begin{align}
T_1(b_1) & = \inf\left\{n-N_0: \left(\max_{n_0\leq t\leq n-n_0} Z_{|\by}(t,n) \right) >b_1,\ n\geq N_0 \right\}, \label{eq:T1} \\
T_2(b_2) & = \inf\left\{n-N_0: \left(\max_{n-n_1\leq t\leq n-n_0} Z_{|\by}(t,n)\right) >b_2,\ n\geq N_0 \right\}, \label{eq:T2} \\
T_3(b_3) & = \inf\left\{n-N_0: \left(\max_{n-n_1\leq t\leq n-n_0} Z_{L|\by}(t,n)\right) >b_3, \ n\geq N_0 \right\}. \label{eq:T3}
\end{align}
Here, $b_1$, $b_2$ and $b_3$ are chosen so that the false discovery rate for each of the stopping rule is controlled at a pre-specified level.  

In the above stopping rules, $n_0, n_1$ and $L$ are pre-specified values.  Usually, $n_0$ is set to be small so as to detect the change as soon as possible, while not too small, such as 1, to avoid the high fluctuations at the very ends.  So $T_1$ is a straightforward stopping rule.  Sometimes, when $\tau$ is large, we may not want to put too much emphasizes on the early observations.  This leads to $T_2$ and $T_3$.   It is easy to see that $T_2$ is a more relaxed version of $T_1$.  In $T_2$, if we set $n_1$ to be $n-n_0$, then it is the same as $T_1$, while we could set $n_1$ tactically to achieve  performance similar to (or even better than) $T_1$ and at the same reduce computation time.

For both $T_1$ and $T_2$, at time $n$, we find $k$ NNs among the first $n$ observations.  One modification we can make is that we use the most recent observations to compute the test statistic.  In $T_3$,  $Z_{L|\by}(t,n)$ is defined the same as $Z_{|\by}(t,n)$ but only based on the $L$ most recent observations:
$\bY_{n-L+1}$, $\dots$, $\bY_n$.  That is, for $i,j\in n_L \overset{\Delta}{=}\{n-L+1,\dots,n\}$, we let
\allowdisplaybreaks
\begin{align*}
A_{n_L,ij}^{(r)} & = \bI(\bY_j \text{ is the $r$th NN of } \bY_i \text{ among observations } \bY_{n-L+1},\dots\bY_n), 
\end{align*}
$A_{n_L,ij}^+  = \sum_{r=1}^k A_{n_L,ij}^{(r)}$, and $R_{L}(t,n)  = \sum_{i,j\in n_L} (A_{n_L,ij}^+ + A_{n_L,ji}^+) B_{ij}(t,n_L)$ with
$B_{ij}(t,n_L) = b_{\mathbf{P}_{n_L}(i)\mathbf{P}_{n_L}(j)}(t)$, where $\mathbf{P}_{n_L}(\cdot)$ is a random permutation among indices $\{n-L+1,\dots,n\}$.  Then
$$ Z_{L|\by}(t,n) = -\frac{R_{L}(t,n) - \ep(R_{L}(t,n))}{\sqrt{\Var(R_{L}(t,n)|\by)}}.$$

\subsection{Comparisons of the three stopping rules}
\label{sec:performance}

Two key objectives of sequential detection are (i) to detect the change-point as soon as possible when it occurs; and (ii) to keep the false discovery rate low.  These can be characterized by two quantities: The expected detection delay, $\ep_{\tau^*} (T-\tau^*|T>\tau^*)$, where $\tau^* = \tau-N_0$ is the time index of the change-point if we set the time we begin the test to be 1; and the average run length, $\ep_\infty(T)$, the expectation of $T$ when there is no change-point or the change-point is at infinity.

In the following, we use Monte Carlo simulations to better understand the three stopping rules.  To make a fair comparison, the critical values $b_i, i=1,2,3$ are chosen (through simulation runs) so that $\ep_\infty(T_i)=2,000$ for each stopping rule.  We then compare their detection delays.
The detailed simulation setup is as follows:  There are $N_0=200$ historical observations from the same distribution.  We begin our test from $t=201$.  In the simulation, the change-point is at $\tau$.  Before the change-point $\tau$, the distribution is a $d$-dimensional Gaussian distribution with mean $\mu_1$ and covariance matrix $I_d$, $\mathcal{N}_d(\mu_1,I_d)$; after the change, the distribution is $\mathcal{N}_d(\mu_2,I_d)$.  Let $\|\mu_2-\mu_1\|_2=2$ where $\|\cdot\|_2$ is the $L_2$ norm.  We consider $\tau=201$ (the change occurs right at the time when we begin to perform the test, i.e., $\tau^*=1$) till $\tau=2201$ (the change occurs 2,000 observations after we begin to perform the test, i.e., $\tau^*=2001$) for an increment of 500.   We consider 1-NN and 3-NN graphs.

\begin{figure}[!htp]
\caption{Boxplots of detection delays of the three stopping rules based on 1,000 simulation runs for each $\tau^*$.  Top panel: $k=1$; bottom panel: $k=3$.  Other parameters are set as: $n_0 = 3, n_1 = 197$, and $L=200$.  The horizontal line is the median of the detection delays for $T_3$ across all 5,000 simulation runs. } \label{fig:delay}
\includegraphics[width=\textwidth]{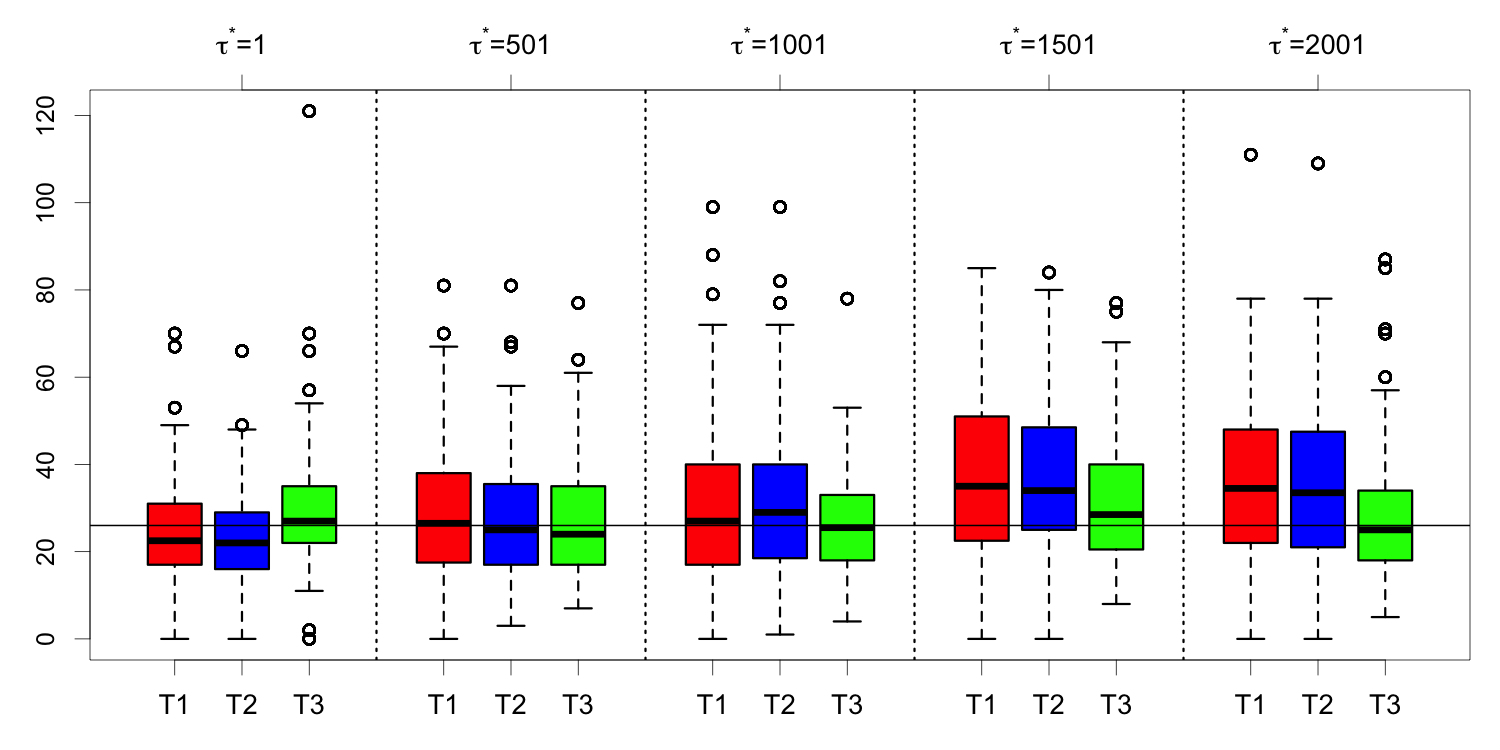}
\includegraphics[width=\textwidth]{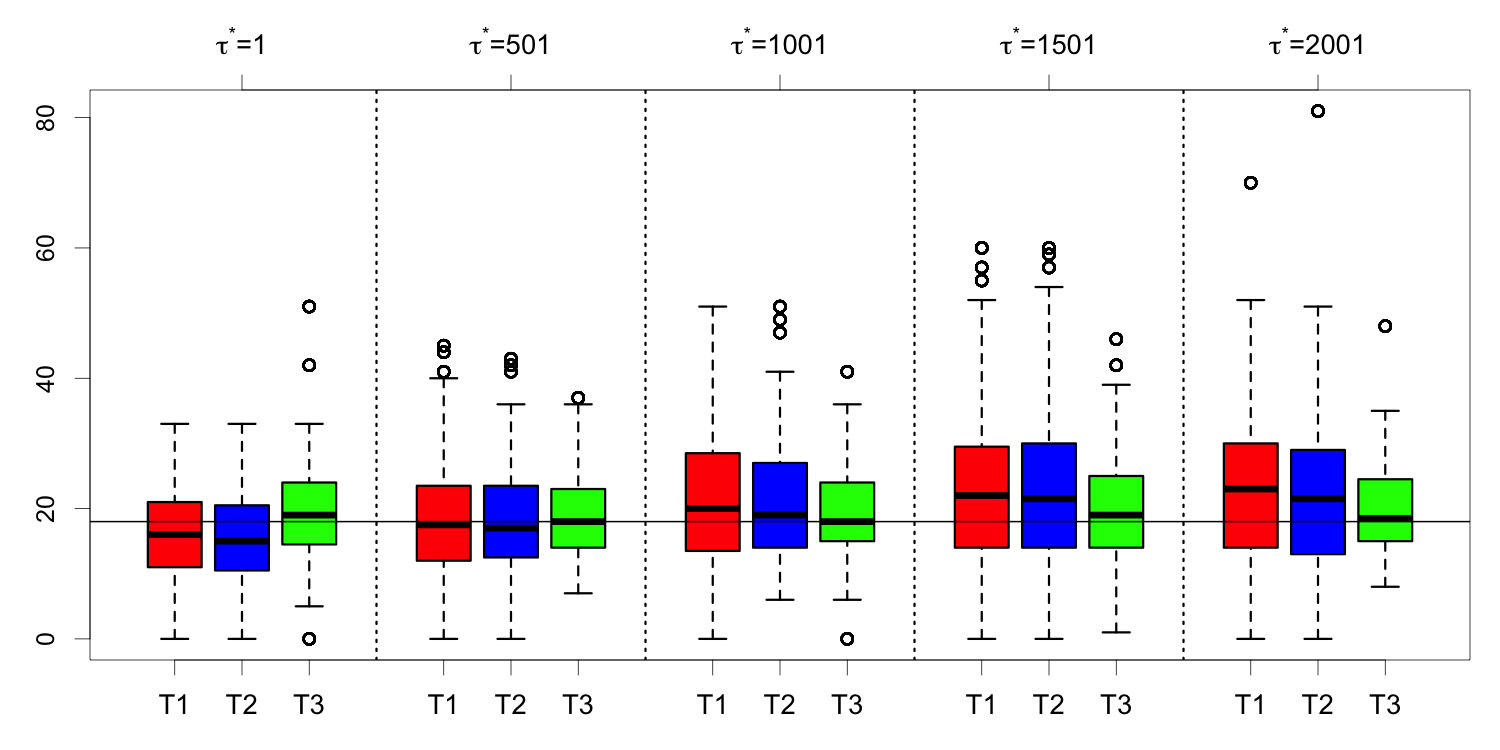}
\end{figure}

Figure \ref{fig:delay} shows boxplots of the detection delays $(T-\tau^*)$ of the three stopping rules under different $\tau^*$'s.   Here, we aim for shorter detection delays.   We can see that $T_2$ is in general slightly better than $T_1$ as the boxes are shifted downward a little bit overall.  When $\tau^*$ is small, $T_3$ has a longer detection delay than $T_1$ or $T_2$.  As $\tau^*$ increases, the detection delay for $T_3$ is almost the same, while that for $T_1$ or $T_2$ increases substantially.  When $\tau^*=1501$, the detection delay for $T_1$ or $T_2$ is clearly larger than $T_3$.  One reason for the increasing detection delay for $T_1$ or $T_2$ is that $Z_{|\by}(t,n)$ is left skewed when the ratio $t/n$ is small and this problem becomes severer as $n$ increases.  

On the other hand, since $T_3$ is based on the same number of observations for all $n$, its detection delay is not affected by where $\tau^*$ locates.  Its detection delay is longer than $T_1$ and $T_2$ when the change occurs at a very early stage, but it is on par with $T_1$ or $T_2$ when the change occurs later, and shorter than $T_1$ and $T_2$ when the change occurs in a late stage.  As the first work on sequential detection based on $k$-NN graphs, we recommend to use $T_3$.  For $T_1$ and $T_2$, one way to overcome the problem of increasing detection delay is to make the thresholds in $T_1$ and $T_2$ to be functions of $n$; for example, we could consider $T_1(b_1(n))$ and $T_2(b_2(n))$ with $b_1(n)$ and $b_2(n)$ monotone increasing functions in $n$.   This is, however, a large topic, and we reserve it for future studies.  

In the following, if not further noted, $T$ and $b$ refer to $T_3$ and $b_3$, respectively.

\section{Average run length $\ep_\infty(T(b))$}
\label{sec:ARL}
Given the stopping rule $T(b)$, the remaining question is how to determine the detection threshold $b$, in particular, how to choose $b$ so that the average run length $\ep_\infty(T(b))$ is a pre-specified value, such as 10,000.  

First of all, we usually don't know the underlying distribution of the observations, so we couldn't directly simulate observations to obtain $b$ as done in Section \ref{sec:performance}.  Secondly, resampling based methods, such as permutation and bootstrap, are not appropriate here as new observations keep arriving and the limited existing observations are usually not representative enough, especially for complicated data.  Even if one could come up with some approaches through resampling methods, they would be very time consuming and not practical for online applications.  Therefore, we seek to obtain an analytic formula for $\ep_\infty(T(b))$.  

Given the non-parametric nature of the proposed method, we would not be able to get an exact analytic formula for $\ep_\infty(T(b))$ for finite $L$, the number of observations used at each time, so we approach the problem asymptotically, i.e., $L\rightarrow \infty$.  We then make further modifications so that the analytic formula is a good approximation for finite $L$.


\subsection{Asymptotic results}
We first consider the asymptotic scenario, $L\rightarrow\infty$.
In this context, $\{Z_{L|\by}(t,n)\}_{t,n}$, with $t$ and $n$ rescaled by $L$, can be shown to converge to a two-dimensional Gaussian random field under very mild conditions.  The properties of the supremum of a two-dimensional Gaussian random field was well studied \citep{siegmund1995using}, and the remaining task is to quantify the covariance function of the Gaussian random field, as well as its partial derivatives.  They can be obtained by studying the dynamics of the NN relations.  The main results are given in Lemma \ref{lemma:Zt} and Theorems \ref{thm:gaussian} and \ref{thm:h12}.

We assume the following condition.
\begin{condition}\label{condition:degree}
There is a positive constant $\mathbb{C},\ 1\leq \mathbb{C}<\infty$, depending only on $k$, such that 
\begin{align*}
\sup_{1\leq j\leq n} \left(\sum_{i=1}^n A_{n,ij}^+ \right)\leq \mathbb{C}, \quad  n\in \mathbb{N}.
\end{align*} 
\end{condition}
In $k$-NN, each observation points to its first $k$ NNs, so the out-degree of each observation (the number of arrows pointing from the observation) is $k$, while the in-degree of each observation (the number of arrows pointing to the observation) can vary.  This condition says that the in-degree of each observation is bounded.  It is satisfied almost surely for multivariate data \citep{bickel1983sums,henze1988multivariate}.  For non-Euclidean data, if the distance is chosen properly, this condition is also easy to hold as many non-Euclidean data can be embedded into a Euclidean space.  

Before stating the main results, we define some useful quantities. According to Propositions 3.1 and 3.2 in \cite{henze1988multivariate}, under Condition \ref{condition:degree}, the quantities $$\frac{1}{L}\sum_{i,j\in n_L}A_{n_L,ij}^{(r)} A_{n_L,ji}^{(s)}, \quad \frac{1}{L}\sum_{i,j,l\in n_L,~j\neq l}A_{n_L,ji}^{(r)} A_{n_L,li}^{(s)},$$ converge in probability to constants as $L\rightarrow\infty$ and the limits can be calculated through complicated integrals \citep{henze1988multivariate}.  We denote the limits as 
\begin{align}
p_{\infty}(r,s) & = \lim_{L\rightarrow\infty} \frac{1}{L}\sum_{i,j\in n_L}A_{n_L,ij}^{(r)} A_{n_L,ji}^{(s)}, \label{eq:q1infty} \\
q_{\infty}(r,s) & = \lim_{L\rightarrow\infty} \frac{1}{L}\sum_{i,j,l\in n_L,~j\neq l}A_{n_L,ji}^{(r)} A_{n_L,li}^{(s)}. \label{eq:q2infty}
\end{align}
Let 
\begin{align}
p_{k,\infty} & = \sum_{r=1}^k\sum_{s=1}^k p_{\infty}(r,s), \\
q_{k,\infty} & = \sum_{r=1}^k\sum_{s=1}^k q_{\infty}(r,s).
\end{align}
Then $p_{k,\infty}$ is the limiting expected number of mutual NNs a node has in $k$-NN and $q_{k,\infty}$ the limiting expected number of nodes that share a NN with a node in $k$-NN.  We also define their finite sample versions by taking expectations:
\begin{align}
p_{L}(r,s) & =  \frac{1}{L}\,\ep\left(\sum_{i,j\in n_L}A_{n_L,ij}^{(r)} A_{n_L,ji}^{(s)}\right),  \\
q_{L}(r,s) & =  \frac{1}{L}\,\ep\left(\sum_{i,j,l\in n_L,~j\neq l}A_{n_L,ji}^{(r)} A_{n_L,li}^{(s)}\right), \\
p_{k,L} & =  \frac{1}{L}\,\ep\left(\sum_{i,j\in n_L}A_{n_L,ij}^+ A_{n_L,ji}^+\right), \label{eq:q1L} \\
q_{k,L} & =  \frac{1}{L}\,\ep\left(\sum_{i,j,l\in n_L,~j\neq l}A_{n_L,ji}^+ A_{n_L,li}^+\right). \label{eq:q2L}
\end{align}
Then $$\lim_{L\rightarrow\infty} p_{L}(r,s) = p_{\infty}(r,s), \ \lim_{L\rightarrow\infty} q_{L}(r,s) = q_{\infty}(r,s),$$
$$ \ \lim_{L\rightarrow\infty} p_{k,L}= p_{k,\infty}, \ \lim_{L\rightarrow\infty} q_{k,L}= q_{k,\infty}.$$  

%

We next state the main results.

\begin{lemma}\label{lemma:Zt}
Under Condition \ref{condition:degree}, when $t-(n-L),\ (n-t)=O(L)$, as $L\rightarrow\infty$, $Z_{L|\by}(t,n)\rightarrow Z_{L}(t,n)$ almost surely, where
\begin{align*}
Z_L(t,n) = -\frac{R_L(t,n)-\ep(R_L(t,n))}{\sqrt{\Var(R_L(t,n))}}.
\end{align*}
\end{lemma}
This lemma follows immediately from Propositions 3.1 and 3.2 in \cite{henze1988multivariate}.  

\begin{theorem}\label{thm:gaussian} 
Under Condition \ref{condition:degree}, as $L\rightarrow\infty$, the finite dimensional distributions of $\{Z_L([vL], [wL]): 0<w-1<v<w<\infty\}$ weakly converges to the finite dimensional distributions of a two-dimensional Gaussian random field, which we denote as $\{Z^\star(v,w):0<w-1<v<w<\infty\}$. (Here, $[x]$ denotes the largest integer smaller than or equal to $x$ for any real number $x$.)
\end{theorem}

A main challenge to prove this theorem is how to deal with the holistic dependencies among $A_{n_L,ij}^+$'s.  
Even for different $i,j,l,r$, $A_{n_L,ij}^+$ and $A_{n_L,lr}^+$ are dependent.  This is because of the constraints $\sum_j A_{n_L,ij}^+=k$ for all $i\in n_L$ (see details in Appendix \ref{sec:GaussianProof}).  

We consider a similar set of Bernoulli random variables $\{\widetilde{A}_{n_L,ij}^+\}_{i,j\in n_L}$ but with relaxed dependencies. 
We keep the following probabilities unchanged:
\begin{eqnarray*}
\pr(\widetilde{A}_{n_L,ij}^+=1) &=& \pr(A_{n_L,ij}^+=1), \\
\pr(\widetilde{A}_{n_L,ij}^+ = 1,\ \widetilde{A}_{n_L,ji}^+ = 1) &=& \pr(A_{n_L,ij}^+ = 1,\ A_{n_L,ji}^+ = 1), \\
\pr(\widetilde{A}_{n_L,ji}^+ = 1,\ \widetilde{A}_{n_L,li}^+ = 1) &=& \pr(A_{n_L,ji}^+ = 1,\ A_{n_L,li}^+ = 1).
\end{eqnarray*}
That is, two-way NN relations are retained.  
However, we relax the other dependencies.
We let $\widetilde{A}_{n_L,ij}^+$ be independent of  $\{\widetilde{A}_{n_L,il}^+, \widetilde{A}_{n_L,li}^+\}_{l\neq j}$.  We also let $\widetilde{A}_{n_L,ij}^+$ and $\widetilde{A}_{n_L,lr}^+$ be independent when $i,j,l,r$ are all different.  

Then $\widetilde{A}_{n_L,ij}^+$'s are only locally dependent.  But $\sum_j \widetilde{A}_{n_L,ij}^+$'s are no longer necessarily $k$. However, $\{\widetilde{A}_{n_L,ij}^+\}_{i,j\in n_L}$ becomes $\{A_{n_L,ij}^+\}_{i,j\in n_L}$ if we condition on the events $\left\{\sum_j \widetilde{A}_{n_L,ij}^+=k\right\}_{i\in n_L}$.  Thus, $Z_L(t,n)$ can be studied through the joint distribution of summations of locally dependent terms.  We then use Stein's method to deal with local dependencies.  The complete proof is in Appendix \ref{sec:GaussianProof}.  

\begin{remark}
The tightness of the two-dimensional field can be shown for $\{Z_L([vL], [wL]): 0<w-1+\delta\leq v\leq w-\delta<\infty\}$ for any $\delta\in (0,1)$.  For $v$ too close to $w-1$ or $w$, the fluctuation in the random field could be too wild to have the field being uniformly tight.  
\end{remark}

Based on Theorem \ref{thm:gaussian}, we approximate $\ep_\infty(T(b))$ by that of the corresponding quantity defined for the limiting random field:
\begin{align}\label{eq:Tstar}
T^\star(b) = \inf\left\{n-N_0: \left(\max_{n-n_1\leq t\leq n-n_0} Z^\star(t/L, n/L)\right)>b,\ n\geq N_0\right\}.
\end{align}
According to \cite{siegmund1995using}, when $b, L, n_0, n_1\rightarrow\infty$ in such a way that $b = c\sqrt{L}$ for some fixed $0<c<\infty$, $n_0=u_0 L$ and $n_1=u_1 L$ for some fixed $0<u_0<u_1<1$, and when there is no change-point, $T^\star(b)$ is asymptotically exponentially distributed with mean
\begin{align}\label{eq:ARL1}
\ep_\infty(T^\star(b)) \sim \frac{\sqrt{2\pi} \exp(b^2/2)}{ c^2\,b \int_{u_0}^{u_1} g_1(u)g_2(u)\nu\left(c\sqrt{2g_1(u)} \right)\nu\left(c\sqrt{2g_2(u)} \right)du},
\end{align}
where
\begin{align*}
g_1(u) & = \left.\frac{\partial_- \rho^\star_{(u,w)}(\delta_1,0)}{\partial \delta_1}\right|_{\delta_1=0} \equiv \left.-\frac{\partial_+ \rho^\star_{(u,w)}(\delta_1,0)}{\partial \delta_1}\right|_{\delta_1=0}, \\
 g_2(u) & = \left.\frac{\partial_- \rho^\star_{(u,w)}(0,\delta_2)}{\partial \delta_2}\right|_{\delta_2=0} \equiv \left.-\frac{\partial_+ \rho^\star_{(u,w)}(0,\delta_2)}{\partial \delta_2}\right|_{\delta_2=0}, \\
\nu(x) & = 2x^{-2} \exp\left\{ -2 \sum_{m=1}^\infty m^{-1} \Phi\left(-\frac{1}{2}xm^{1/2} \right)\right\}, \quad x>0.
\end{align*} 
Here, $\rho^\star_{(u,w)}(\delta_1,\delta_2) = \Cov(Z^\star (w-u,w), Z^\star (w-u+\delta_1, w+\delta_2))$ and
$\nu(\cdot)$ is closely related to the Laplace transform of the overshoot over the boundary of a random walk.  A simple approximation given in \cite{siegmund2007statistics} is sufficient for numerical purpose:
\begin{equation}
  \label{eq:nu_approx}
  \nu(x) \approx \frac{(2/x)(\Phi(x/2)-0.5)}{(x/2)\Phi(x/2)+\phi(x/2)},
\end{equation}
where $\Phi(\cdot)$ is the cumulate distribution function of the standard normal distribution and $\phi(\cdot)$ the density function of the standard normal distribution.

Thus, the remaining task is to derive the directional partial derivatives of the covariance function of the Gaussian random field.  Their analytic expressions are given in the following theorem.

\begin{theorem}\label{thm:h12}
For the two-dimensional field $\{Z^\star(v,w):0<w-1<v<w<\infty\}$, the directional partial derivatives are
\begin{align}
g_1(u) & = \left. \frac{\partial_- \rho^\star_{(u,w)}(\delta_1,0)}{\partial \delta_1} \right|_{\delta_1=0} \equiv \left. -\frac{\partial_+ \rho^\star_{(u,w)}(\delta_1,0)}{\partial \delta_1}\right|_{\delta_1=0}  \label{eq:h1} \\
& = \frac{16u(1-u)(k+p_{k,\infty})+2(1-2u)^2(q_{k,\infty}-k^2+k)}{\sigma^2(u)}, \nonumber
\end{align}
\begin{align}
& g_2(u)  = \left. \frac{\partial_- \rho^\star_{(u,w)}(0,\delta_2)}{\partial \delta_2} \right|_{\delta_2=0}  \equiv \left. -\frac{\partial_+ \rho^\star_{(u,w)}(0,\delta_2)}{\partial \delta_2} \right|_{\delta_2=0}  \label{eq:h2}  \\
& \ = \frac{16u^2(1-u)^2(p_{k,\infty}+q_{k,\infty}+k^2+2p_{k,\infty}^{(k)}-2q_{k,\infty}^{(k)})}{\sigma^2(u)}  \nonumber \\
& \quad  \ + \frac{4u(1-u)(2q_{k,\infty}^{(k)}-3q_{k,\infty}+k^2+k) + 2(q_{k,\infty}-k^2+k) }{\sigma^2(u)}, \nonumber 
\end{align}
where 
\begin{align*}
& \sigma^2(u)  = 4u(1-u)(4u(1-u)(k+p_{k,\infty}) + (1-2u)^2(q_{k,\infty}-k^2+k)), \\
& p_{k,\infty}^{(k)}  = \sum_{r=1}^k p_{\infty}(k,r), \quad
q_{k,\infty}^{(k)} = \sum_{r=1}^k q_{\infty}(k,r).
\end{align*}
\end{theorem}

The complete proof of this theorem is in Appendix \ref{sec:proofthmh12}.  We studied the dynamics of the $k$-NN series as new observations are added through combinatorial analysis and it turned out that a few key quantities are enough to characterize the dynamics in the asymptotic domain. 

%
%
%
%

\subsection{Finite $L$}
\label{sec:skew}

We now consider the practical scenario where $L$ is finite.  Based on Theorems \ref{thm:gaussian} and \ref{thm:h12}, $\ep_\infty(T(b))$
can be approximated by 
$$\ep_\infty(T(b))\approx \frac{L\sqrt{2\pi} \exp(b^2/2)}{b^3  \int_{\frac{n_0}{L}}^{\frac{n_1}{L}} g_1(u)g_2(u)\nu\left(\sqrt{2b^2g_1(u)/L} \right)\nu\left(\sqrt{2b^2g_2(u)/L} \right)du}$$
with the analytic expressions for $g_1(u)$ and $g_2(u)$ given in \eqref{eq:h1} and \eqref{eq:h2}, respectively, and $\nu(\cdot)$ given in \eqref{eq:nu_approx}.

When deriving the limiting expressions for $g_1(u)$ and $g_2(u)$, we evaluate $\sum_{j} \ep\left(A_{n_L,ij}^{(r)} A_{n_L,ji}^{(s)}\right)$ and $\sum_{j\neq l} \ep\left(A_{n_L,ji}^{(r)} A_{n_L,li}^{(s)}\right)$ under $L\rightarrow\infty$ and the two quantaties become $p_{\infty}(r,s)$ and $q_{\infty}(r,s)$, respectively.  In practice, when $L$ is finite, $p_{\infty}(r,s)$ and $q_{\infty}(r,s)$ are not the best estimates for these two expectations, yet the expectations could be better estimated through historical data.  Therefore, we use the following formula to approximate $\ep_\infty(T(b))$ in practice:
\begin{align}\label{eq:ARL2}
\ep_\infty(T(b))\approx \frac{L\sqrt{2\pi} \exp(b^2/2)}{b^3 \int_{\frac{n_0}{L}}^{\frac{n_1}{L}} g_{L,1}(u)g_{L,2}(u)\nu\left(\sqrt{2b^2 g_{L,1}(u)/L} \right)\nu\left(\sqrt{2b^2 g_{L,2}(u)/L} \right)du}
\end{align}
where $g_{L,1}(u)$ and $g_{L,2}(u)$ are the same as $g_1(u)$ and $g_2(u)$, respectively, except that $p_{k,\infty}$, $q_{k,\infty}$, $p_{k,\infty}^{(k)}$ and $q_{k,\infty}^{(k)}$ are replaced by $p_{k,L}$, $q_{k,L}$, $p_{k,L}^{(k)}$ and $q_{k,L}^{(k)}$, respectively, with $p_{k,L}$ given in \eqref{eq:q1L}, $q_{k,L}$ given in \eqref{eq:q2L}, and
\begin{align}\label{eq:qL}
p_{k,L}^{(k)} & = \sum_{r=1}^k p_{L}(k,r), \quad
q_{k,L}^{(k)}  = \sum_{r=1}^k q_L(k,r).
\end{align}
For $p_{k,L}$, $q_{k,L}$, $p_{k,L}^{(k)}$ and $q_{k,L}^{(k)}$, they usually don't have analytical expressions.  However, they can be easily estimated from historical data.  These estimates can further be updated by new observations as long as no change-point is detected.

We next check how this analytic approximation works.  We compare the threshold $b$ such that $\ep_\infty(T(b))=10,000$ based on this analytic approximation and that based on 10,000 Monte Carlo simulations.  The threshold obtained through 10,000 Monte Carlo simulations can be regarded as the true threshold.  Results under 
different choices of $n_0$, $k$ and $d$ for multivariate Gaussian data are shown in Table \ref{table:pvalue2}.  We checked two values of $L$, namely $L=200$ and $L=50$, and let $n_1=L-n_0$.

\begin{table}[!htp]
\caption{The threshold $b$, such that $\ep_\infty(T(b))=10,000$, through 10,000 Monte Carlo simulations, through analytic formula \eqref{eq:ARL2} based on asymptotic results, and through analytic formula \eqref{eq:ARL3} with additional skewness correction.  Each observation in the data sequence follows a $d$-dimensional normal distribution.} \label{table:pvalue2}
\begin{center}
\begin{tabular}{|c|c||c|c|c|c|c|c|}
\hline
\multicolumn{2}{|c||}{} & \multicolumn{3}{c|}{$n_0=3$}  & \multicolumn{3}{|c|}{$n_0=10$}   \\ \cline{3-8}
 \multicolumn{2}{|c||}{}  & Monte  & Asymp. & Skewness & Monte & Asymp. & Skewness \\
 \multicolumn{2}{|c||}{}  & Carlo  & \eqref{eq:ARL2} & Corrected & Carlo  & \eqref{eq:ARL2} & Corrected \\ 
 \multicolumn{2}{|c||}{} & & & \eqref{eq:ARL3} & & & \eqref{eq:ARL3} \\ \hline \hline
\multicolumn{8}{|l|}{$\mathbf{L=200}$} \\ \hline
\multirow{3}{*}{$d=10$} & $k=1$ & 4.04  & 4.40 & {4.07} & 4.04 & 4.31 & {4.07}   \\ 
& $k=3$ & 4.14 & 4.34 & {4.14}  & 4.14  & 4.23 & {4.14} \\ 
& $k=5$ & 4.16 & 4.31  & {4.18} & 4.16  & 4.17 & {4.18} \\ \hline 
\multirow{2}{*}{$d=100$} & $k=1$ & 3.76  & 4.37 & {3.79} & 3.76  & 4.26 & {3.79} \\ 
& $k=3$ & 3.78  & 4.33 & {3.79} & 3.78  & 4.20 & {3.79} \\ 
& $k=5$ & 3.79 & 4.31 & {3.81} & 3.79  & 4.18 & {3.81} \\ \hline \hline
\multirow{2}{*}{$d=1000$} & $k=1$ & 3.73  & 4.38 & {3.73} & 3.73  & 4.28 & {3.73} \\ 
& $k=3$ & 3.71  & 4.33 & {3.71} & 3.71  & 4.21 & {3.71} \\ 
& $k=5$ & 3.75 & 4.32 & {3.72} & 3.75  & 4.18 & {3.72} \\ \hline \hline
\multirow{2}{*}{$d=10000$} & $k=1$ & 3.71  & 4.38 & {3.70} & 3.71  & 4.27 & {3.70} \\ 
& $k=3$ & 3.65  & 4.33 & {3.69} & 3.65  & 4.21 & {3.69} \\ 
& $k=5$ & 3.68 & 4.32 & {3.69} & 3.68  & 4.18 & {3.69} \\ \hline \hline
\multicolumn{8}{|l|}{$\mathbf{L=50}$} \\ \hline
\multirow{3}{*}{$d=10$} & $k=1$ & 4.00  & 4.38 & {4.10} & 3.99 & 4.24 & {4.10}  \\ 
& $k=3$ & 4.36 & 4.32 & {4.37} & 4.36  & 4.19 & {4.37} \\ 
& $k=5$ & 4.57  & 4.28 & {4.50} & 4.57  & 4.15 & {4.50} \\ \hline 
\multirow{2}{*}{$d=100$} & $k=1$ & 3.86  & 4.36 & {3.94} & 3.83  & 4.23 & {3.94} \\ 
& $k=3$ & 3.92  & 4.31 & {4.02} & 3.92  & 4.18 & {4.02} \\ 
& $k=5$ & 3.95  & 4.29 & {4.09} & 3.95  & 4.15 & {4.09} \\ \hline \hline
\multirow{2}{*}{$d=1000$} & $k=1$ & 3.83  & 4.36 & {3.91} & 3.83  & 4.23 & {3.91} \\ 
& $k=3$ & 3.92  & 4.32 & {3.93} & 3.92  & 4.18 & {3.93} \\ 
& $k=5$ & 3.95  & 4.29 & {3.97} & 3.95  & 4.15 & {3.97} \\ \hline \hline
\multirow{2}{*}{$d=10000$} & $k=1$ & 3.79  & 4.36 & {3.90} & 3.79  & 4.23 & {3.90} \\ 
& $k=3$ & 3.86  & 4.32 & {3.90} & 3.86  & 4.18 & {3.90} \\ 
& $k=5$ & 3.91  & 4.29 & {3.92} & 3.91  & 4.15 & {3.92} \\ \hline \hline
\end{tabular}
\end{center}
\end{table}

Unfortunately, the thresholds obtained through the analytic approximation \eqref{eq:ARL2} are not that close to the Monte Carlo results except for a few occasions.  The analytic approximation \eqref{eq:ARL2} gives similar thresholds for different dimensions when all other parameters are fixed.  However, the thresholds from Monte Carlo simulations are quite different for different dimensions with those under a higher dimension much smaller.  Thus, \eqref{eq:ARL2} is still missing some major components for finite $L$ due to the fact that $Z_L(t,n)$ can be quite left skewed for finite $L$ and small $(n-t)$.  In the following, we incorporate skewness of $Z_L(t,n)$ to improve the analytic approximation.

\begin{remark}
The reason of the discrepancy between the asymptotic results and finite samples was discussed in details in the offline counterpart of the work  \citep{chen2015graph}.  Briefly, the convergence rate of $Z_L(t,n)$ to the Gaussian distribution is slow if $(n-t)/L$ is close to 0 or 1.   In this online detection setting, the problem is even severer as we would like to set $n_0$ very small (such as 3) so as to detect the change as soon as it happens.  For finite $L$, $Z_L(t,n)$ is quiet left skewed when $(n-t)$ is close to 0 or $L$, and the tail probability is overestimated, making the threshold $b$ obtained based on the asymptotic results too conservative.
\end{remark}




\subsubsection{Skewness correction}
\label{sec:skewness}

We adapt the skewness correction approach in \cite{chen2015graph}.  In particular, when we derive the average run length for the limiting two-dimensional Gaussian random field \eqref{eq:ARL1}, the term in the integral related to the marginal distribution of $Z^\star(w-u,w)$ is $\pr(Z^\star(w-u,w)\in b+du)$. (Here, $du$ is the differential of the variable $u$, and similar definition for $dt$ in the following.)  To make the analytic approximation more accurate for finite $L$ and small $n_0$, we replace $\pr(Z^\star(w-u,w)=b+du)$ by an estimate of $\pr(Z_L([n(w-u)],[nw])\in b+du)$.  Following the method based on cumulant-generating functions and change of measure (details refer to \cite{chen2015graph}), we have
\begin{align}\label{eq:Su}
& \frac{\pr(Z_L(t,n)\in b+dt/b)}{\pr(Z^\star(n/L-t/L, n/L)\in b+dt/b)} \\ &\quad \quad \quad \quad \approx \frac{\exp((b-\theta_b)^2/2 + \theta_b^2 \gamma_L(t,n) \theta_b/6}{\sqrt{ (1+\gamma_L(t,n)\theta_b}} := S_L((n-t)/L). \nonumber
\end{align}
Here, $\theta_b = (-1+\sqrt{1+2\gamma_L(t,n) b})/\gamma_L(t,n)$ and $\gamma_L(t,n) = \ep(Z_L^3(t,n))$.
The denotation for $S_L((n-t)/L)$ holds because $\gamma_L(t,n)$ relates to $t$ and $n$ only as a function of $n-t$ (see Lemma \ref{lemma:gamma} below).  Then, the analytic approximation for $\ep_\infty(T)$ incorporating skewness becomes
\begin{align}\label{eq:ARL3}
\frac{L \sqrt{2\pi} \exp(b^2/2)}{b^3 \int_{n_0/L}^{n_1/L}\mathbf{S_L(u)} g_{L,1}(u)g_{L,2}(u)\nu\left(\sqrt{2b^2\, g_{L,1}(u)/L} \right)\nu\left(\sqrt{2b^2\, g_{L,2}(u)/L} \right)du}.
\end{align}
When $L\rightarrow\infty$ and $(n-t), (L-(n-t)) = O(L)$, $\gamma_L(t,n)$ goes to 0 and $S_L(u)$ goes to $1$ for $0<u<1$, so this formula converges to \eqref{eq:ARL1} in the limit.

The exact analytic expression for $\gamma_L(t,n)$ is given in the following lemma:
\begin{lemma}\label{lemma:gamma}
We have 
\begin{align*}
\gamma_L(t,n) = \frac{(\ep(R_L(t,n)))^3 + 3\ep(R_L(t,n)) \Var(R_L(t,n)) - \ep(R_L^3(t,n))}{(\Var(R_L(t,n)))^{3/2}},
\end{align*}
where $\ep(R_L(t,n))$ and $\Var(R_L(t,n))$ are given in \eqref{eq:ERL} and \eqref{eq:VRL}, respectively, and 
\begin{align*}
& \ep(R_L^3(t,n)) = 8k^3L^3\,r_4 + 12k^2L^2(r_2+3k(r_2-2\,r_4)) \\
& + 4kL(3\,r_2-r_1+2\,r_3-4\,r_4 + 3k(3\,r_1-2\,r_2-4\,r_3-4\,r_4)+8k^2(r_3-3\,r_2+5\,r_4)) \\
& + 24\,p_{k,L}(kL^2\,r_4 + kL(r_1+r_2-2r_3-4r_4)+2L(2r_3-r_1+2r_4)) \\
& + 12\,q_{k,L}(kL^2(r_2-2r_4)+kL(2r_3-5r_2+8r_4)+L(r_1+r_2-2r_3-4r_4)) \\
& + 4\,(2\,r_3-3\,r_2+4\,r_4)\,\ep\left(\sum_{i,j,l,v} A_{n_L,ji}^+ A_{n_L,li}^+ A_{n_L,vi}^+\right)  \\
& + 24\,(r_1+r_2-2\,r_3-4\,r_4)\,\ep\left(\sum_{i,j,l}A_{n_L,ij}^+ A_{n_L,ji}^+ A_{n_L,li}^+\right) \\
& + 24\,(2\,r_4-r_2)\,\ep\left(\sum_{i,j,l,v} A_{n_L,ij}^+ A_{n_L,li}^+ A_{n_L,vj}^+ \right) \\
& -16\,r_4 \left(\ep\left(\sum_{i,j,l}A_{n_L,ij}^+ A_{n_L,jl}^+ A_{n_L,li}^+\right) + 3\, \ep\left(\sum_{i,j,l} A_{n_L,ij}^+ A_{n_L,il}^+ A_{n_L,jl}^+ \right) \right) 
\end{align*}
with
\begin{align*}
r_1 & = \frac{2x(L-x)}{L(L-1)}, \quad x=L-(n-t), \\
r_2 & = \frac{4x(x-1)(L-x)(L-x-1)}{L(L-1)(L-2)(L-3)}, \\
r_3 & = \frac{x(L-x)((x-1)(x-2)+(L-x-1)(L-x-2))}{L(L-1)(L-2)(L-3)}, \\ 
r_4 & = \frac{8x(x-1)(x-2)(L-x)(L-x-1)(L-x-2)}{L(L-1)(L-2)(L-3)(L-4)(L-5)}.
\end{align*}
\end{lemma}
To prove this lemma, we have
\begin{align*}
\ep&(R_L^3(t,n)) = \ep(\ep(R_L^3(t,n)|\bY)) \\
& = \sum_{i,j,l,r,u,v}\ep\left((A_{n_L,ij}^++A_{n_L,ji}^+)(A_{n_L,lr}^++A_{n_L,rl}^+)(A_{n_L,uv}^++A_{n_L,vu}^+)\right)\\
& \quad \quad \quad \quad \quad \times \ep\left(B_{ij}(t,n_L)B_{lr}(t,n_L)B_{lr}(t,n_L)\right).
\end{align*}
Adapting similar arguments in calculating the covariance in the proof of Theorem \ref{thm:h12} but with more careful treatment of the summation indices, we could get the result in the lemma.

\begin{figure}[!htp]
\includegraphics[width=\textwidth]{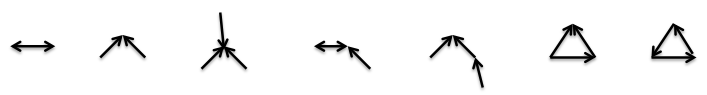}
\caption{The configurations in $k$-NN that relate to the third moment of $R_L(t,n)$.}\label{fig:edge}
\end{figure}

From Lemma \ref{lemma:gamma}, we see that
$\ep(R_L^3(t,n))$ depends on the probability of having certain structures in the nearest neighbor graph.  The relevant structures in $k$-NN are shown in Figure \ref{fig:edge}.  The first two structures represent mutual NNs and shared NNs.  The other five structures are three-way interactions among the NN relations.  The probability of having each of them can be estimated through historical data, and can also be updated by new observations when no change-point is detected. 


We now check how skewness correction performs.  
Table \ref{table:pvalue2} also lists the thresholds obtained through the analytic approximation with skewness correction.
We see that, after skewness correction, the analytic formula gives much better estimates to the thresholds.  When $L=200$, all thresholds estimated by \eqref{eq:ARL3} are very accurate.  Even for small $L$ ($L=50$), the analytic approximated with skewness correction is doing a reasonable job.  When the dimension becomes larger, the threshold estimated by the analytic formula with skewness correction is smaller, exhibiting the same trend as the Monte Carlo results.  

These results show that the formula with skewness correction could capture the major factors and gives quite reliable estimates.  It would be reasonable to use the analytic formula with skewness correction to get the threshold $b$ in real applications.

\section{Power analysis}
\label{sec:EDD}

Given the procedure and the fast analytic way of determining the detection threshold, the proposed method can be easily applied to real problems.  Now, the question is how powerful this method is.  To get some idea, we compare it to the test based on Hotelling's $T^2$ test for multivariate Gaussian data as Hotelling $T^2$ test is asymptotically the most powerful for testing two multivariate Gaussian distributions with the same covariance matrix.

The simulation setup is as follows: There are $N_0=200$ historical observations and a change occurs at $t=400$ (200 new observations after the start of the test).  The observations are independent and follow $d$-dimensional Gaussian distribution with a mean shift ($\Delta$) at the change-point. (The $L_2$ distance between the two means is $\Delta$.)
The amount of change, $\Delta$, is chosen so that the tests have moderate power.  Results are given in Table \ref{table:power}.  ``Successful detection" is defined the same as in Section 
\ref{sec:performance} that the test detects the change-point within 100 observations after the change occurred.  We compare all tests on the same ground by controlling the early stop probability to be 0.01. 


\begin{table}[!htp]
\caption{Fraction of trials (out of 1,000) that the change-point is successfully detected for the proposed test and for the test based on the Hotelling's $T^2$ test.} 
\label{table:power}
\begin{center}

 
%

\begin{tabular}{|c||c|c|c|c|c|c|}
\hline
& \multicolumn{4}{c}{Normal data} & \multicolumn{2}{|c|}{Log-normal data} \\ \cline{2-7}
& $d=10$ & $d=100$ & $d=1000$ & $d=10000$ & $d=10$ & $d=100$  \\
& $ \Delta=0.7$ & $ \Delta=1.8$ & $\Delta=2.7 $ & $\Delta=5 $ & $\Delta=1.5$ & $\Delta=2$   \\ \hline \hline
 Proposed test: 1-NN & 0.02 & 0.21& 0.12 & 0.16 &  0.48 & 0.08  \\ \hline
 Proposed test: 3-NN & 0.07 & 0.55 & 0.41 & 0.52 &  0.87 & 0.48 \\ \hline
 Proposed test: 5-NN & 0.15 & \textbf{0.81} & \textbf{0.57} & \textbf{0.70} &  \textbf{0.95} & \textbf{0.77}  \\ \hline
  Hotelling's $T^2$ & \textbf{0.69} &  0.63 & -- & -- & 0.34 & 0.02  \\ \hline \hline
\end{tabular}

\end{center}
\end{table}

Table \ref{table:power} shows the results under different scenarios with 1,000 simulation runs for each scenario.  The fraction of the runs that the change-point is successfully detected is reported.  
When the data is multivariate Gaussian, we see that the test based on the Hotelling $T^2$ test is doing very well in low dimension.  When the dimension becomes higher, the power of the proposed test catches up.  When $d=100$, the proposed test based on 5-NN is outperforming the test based on the Hotelling $T^2$ test.  When $d$ is even higher, the dimension is larger than the number of observations that the method based on the Hotelling's $T^2$ cannot be applied.  For the proposed tests, we see that the we do need to increase the strength of the signal to achieve a similar detection power.  However, the number of fold we need for the increase of the signal is much smaller than that for the dimension.  When the dimension increase from 100 to 10000 (by a fold of 100), we only need to increase the signal by a fold about 3 to achieve the same detection power.  Hence, the proposed method is relatively mildly affected by the dimensionality.

We also did the comparison for log-normal data and the change is in the mean parameter.  Now, the assumptions for the Hotelling $T^2$ test do not hold and we see that the proposed test is outperforming the test based on the Hotelling $T^2$ test even when the dimension is low.

The results show that the proposed test has satisfying power and works for various distributions.

\section{An illustration example from real data}
\label{sec:realdata}

Here, we apply the proposed method to a real dataset on network analysis.  The dataset has been completely collected at the time of analysis.  We treat it as if the data were being observed to illustrate how the proposed method works.  It is conceivable to apply the proposed method in a sequential manner if the data keep arriving.

The MIT Media Laboratory conducted a study following 106 subjects, students and stuff in an institute, who used mobile phones with pre-installed software that can record all activities on their phones from July 2004 to June 2005 \citep{eagle2009inferring}.  A natural question of interest is whether there is any change in the phone-call pattern among these people over time.  This is one way to assess their friendship along time.  

We bin the phone calls by day, and for each day, construct a phone-call network with the subjects as nodes and a directed edge pointing from subject $i$ to subject $j$ if subject $i$ called subject $j$ on that day.  
We encode the directed network of each day by an adjacency matrix, with 1 for element $[i,j]$ if there is a directed edge pointing from subject $i$ to subject $j$, and 0 otherwise.  Let $M_i$ be the $106 \times 106$ adjacency matrix on day $i$.  We consider two distance measures defined as:
\begin{enumerate}[(1)]
\item the number of different entries: $\|M_i-M_j\|_F^2$, where $\|\cdot\|_F$ means the Frobenius norm of a matrix,
\item the number of different entries, normalized by the geometric mean of the total edges in each day: $\frac{\|M_i-M_j\|_F^2}{\|M_i\|_F \|M_j\|_F}$.
\end{enumerate}

\begin{figure}[!htp]
\begin{center}
\includegraphics[width=.8\textwidth]{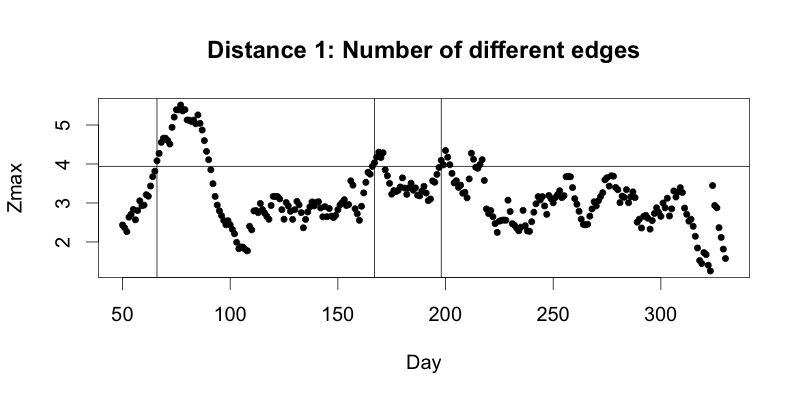}
\includegraphics[width=.8\textwidth]{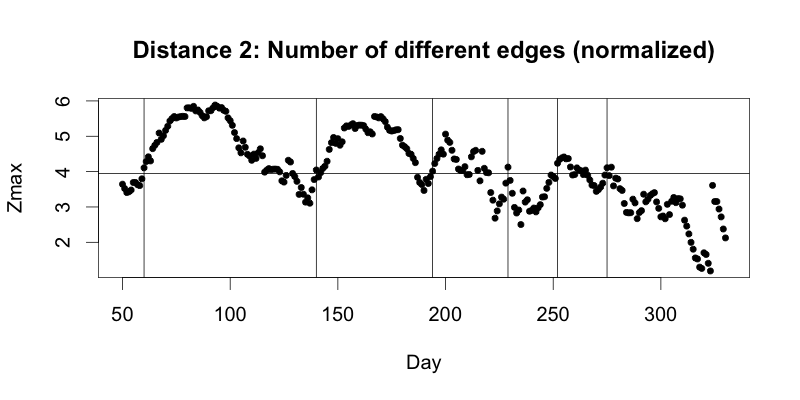}
\end{center}
\caption{Zmax for the network data based on two distances.  The horizontal line in each plot is the threshold $b$ such that $\ep_\infty(T(b))=10,000$.  The vertical lines are the valid stopping times.} \label{fig:networkD12}
\end{figure}


For this dataset, since there is no further information to tell whether there is any change-point for the first few observations, we applied the offline change-point detection method in \cite{chen2015graph} on the first 50 days/observations.  No change-point was found for either distance measure.  So we treat the first 50 observations as historical observations.  We let $L=50$, $n_0=3$, and determine the threshold based on \eqref{eq:ARL3}.

Figure \ref{fig:networkD12}  plots $\text{Zmax}(n) = \max_{n-L+n_0\leq t\leq n-n_0} Z_{L|\by}(t,n)$ against $n$, the index of days, based on the two distances.  The detection thresholds for the two distances are $b=3.92$ and $b=3.98$, respectively.  Since multiple stopping times might be called for one change-point, we disregard time $n$ if $\max(\text{Zmax}(n-5), \text{Zmax}(n-4), \dots, \text{Zmax}(n-1))>b$, i.e., we consider them to be caused by the same event.  We call the remaining stopping times the ``candidate stopping times".
Then, 
three candidate stopping times for distance 1 and six candidate stopping times for distance 2 are found.  They are summarized in Table \ref{table:networkEvents}, together with their nearby academic events.

\begin{table}[!htp]
\caption{Valid stopping times and their nearby academic events.} 
\label{table:networkEvents}
\begin{center}
\begin{tabular}{|c|l|l|}
\hline
Distance 1 & \multicolumn{1}{c}{Distance 2} & \multicolumn{1}{|c|}{Nearby academic event*} \\ \hline
$n=66$: 2004/9/23 & $n=60$: 2004/9/17 & 9/9: First day of class for Fall term \\ \hline
$n=167$: 2005/1/2 & $n=140$: 2004/12/6 & 12/18: Last day of class for Fall term \\ \hline
$n=198$: 2005/2/2 & $n=194$: 2005/1/29 & 2/2: First day of class for Spring term \\ \hline
--- & $n=229$: 2005/3/5 &  3/5: Registration deadline for Spring term  \\ \hline
--- & $n=252$: 2005/3/28 & 3/21: Spring vacation \\ \hline
--- & $n=275$: 2005/4/20 & 4/21: Drop deadline for Spring term \\ \hline
\end{tabular}
\end{center}
* The dates of the academic events are from the 2015-2016 academic calendar of MIT as the 2004-2005 academic calendar of MIT cannot be found online.
\end{table}

From Table \ref{table:networkEvents}, we see that the proposed method based on either distance finds change-points at around the beginning of the Fall term, the end of the Fall term, and the beginning of the Spring term.  The proposed method using distance 2 finds additional change-points in the middle of the Spring term.  These are all reasonable times to have some significant call pattern changes.

\begin{figure}[!htp]
\begin{center}
\vspace{-1em}
\includegraphics[width=.9\textwidth]{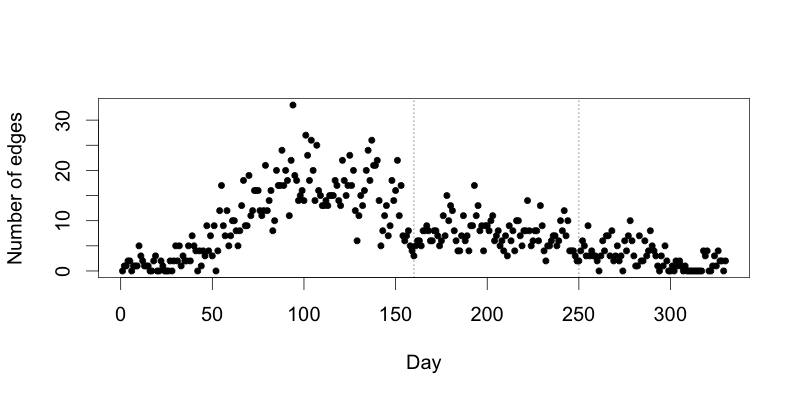}
\end{center}
\vspace{-2em}
\caption{Number of edges in the phone-call network for each day.} \label{fig:dayCount} 
\end{figure}

One may wonder if these change-points could be found by a 1-dimensional summary statistic.  We plot in Figure \ref{fig:dayCount} the number of edges in each network over time.  We could see clearly the change-points at around the beginning of the Fall term and the end of the Fall term, reflected by the change of the call volume.  Starting from the winter break ($n=160$), the call volume stabilizes.  There is a slight call volume decrease starting from the spring vacation (at around $n=250$).  However, the call volumes from $n=160$ toward $n=250$ are quite similar, and there is no significant change within this period.  For example, we apply the function \emph{cpt.meanvar()} in \texttt{R} package \texttt{changepoint}, a 1-dimensional change-point detection approach for detecting either mean or variance change, to this segment of data and no change-point is found.  Hence, there is no significant change in the call volume transiting from the winter break to the Spring term.

On the other hand, the proposed method on either distance finds the change-point at the beginning of Spring term (around $n=198$), indicating that there are some structural changes in the phone-call network which wouldn't be captured by only examining the call volume.  Also, since distance 2 is normalized by the total number of edges in each network, there are probably structural changes in the phone-call network besides call volume change in the other five change-points detected based on distance 2.  

For further details on what the changes are, one could pick some networks before and after the change-point and conduct more detailed comparisons.  Moreover, if one is interested in some specific characteristics of the network, a distance reflecting such characteristics can be used for the proposed method.

\begin{remark}
This phone-call network example is for illustration.  The proposed method will be more useful in real data applications where data are indeed being collected (not completely collected at the time of the analysis) and the observations cannot be characterized through a simple model, such as 
a long vector with unknown structures among the elements, a combination of quantitative and qualitative components, a networks, or an image, with the type of change not specified.
\end{remark}



\section{Dicussion}
\label{sec:discussion}

In the section, we briefly discuss the choice of $k$, the number of NNs to be included in the test, how the test works for gradual changes, and possible extensions of the tests to other graphs.

\subsection{Choice of $k$}
\label{sec:kchoice}



Heuristically, if we choose a very small value of $k$, some useful similarity information among the observations is not used by the test.  We see from Table \ref{table:power} that the power of the test increases from 1-NN to 5-NN .  On the other hand, if we set $k$ to be too large, it may include some irrelevant information, which would also harm the power.

\begin{figure}[!htp]
\includegraphics[width=0.85\textwidth]{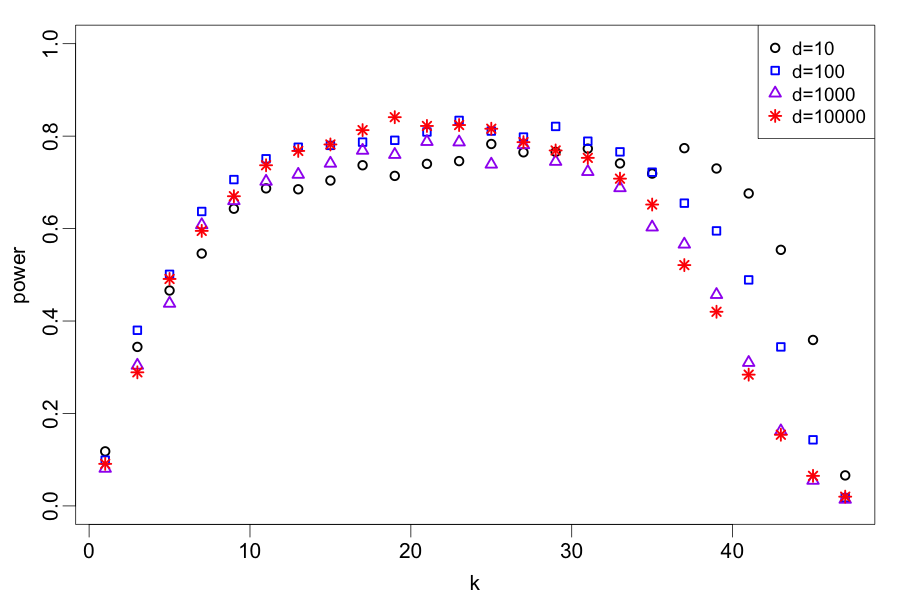}
\caption{Power of the test based on $k$-NN for detecting change-points in sequences of multivariate normal data over a range of dimensions: $d=10$ (black circle), $d=100$ (blue square), $d=1000$ (purple triangle), and $d=10000$ (red star).  The change is a shift in mean with the $L_2$ distance between the means before and after the change $\Delta$: $\Delta=1.7$ ($d=10$), $\Delta=2.7$ ($d=100$), $\Delta=4.5$ ($d=1000$), and $\Delta=8$ ($d=10000$).  The parameter $L$ is set to be 50, and the power is estimated through 1000 simulation runs.}\label{fig:kchoice50}
\end{figure}

\begin{figure}[!htp]
\includegraphics[width=0.85\textwidth]{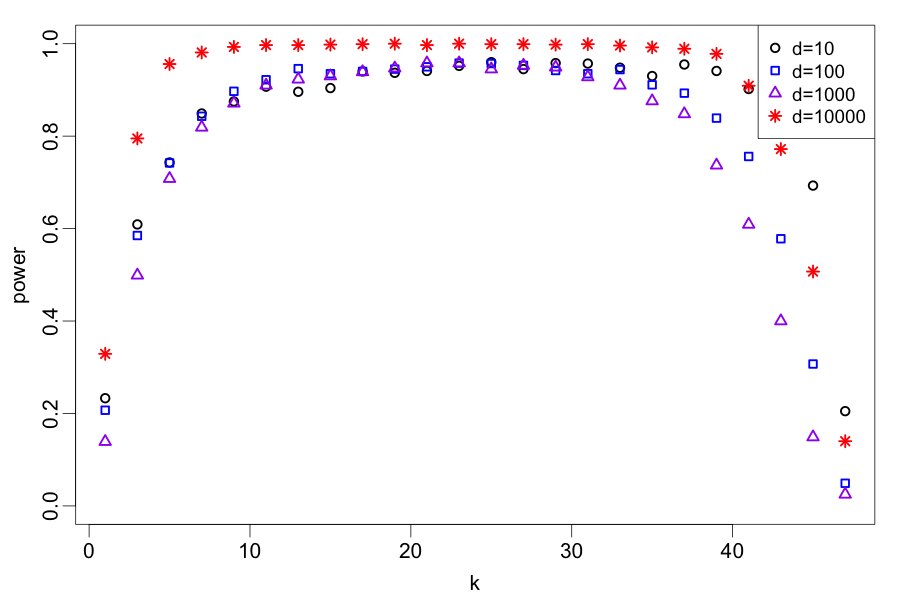}
\caption{The same set up as in Figure \ref{fig:kchoice50}  while the strength of the signal is  increased for each dimension: $\Delta=2$ ($d=10$), $\Delta=3$ ($d=100$), $\Delta=5$ ($d=1000$), and $\Delta=10$ ($d=10000$).}\label{fig:kchoice50_2}
\end{figure}

Figure \ref{fig:kchoice50} plots the power of the test as $k$ varies.  The different symbols corresponds to different dimensions of the observations.  For each dimension, the amount of the change is fixed and only $k$ varies.  The amount of the change for each dimension is chosen so that the highest power is around 0.8.  We can see clearly from the plot the relation between the power and $k$: The power first increases as $k$ increases and becomes steady for a wide range of $k$'s and then decreases as $k$ increases.  Therefore, the optimal $k$ should be chosen before the test reaches the plateau to achieve a high power and low computation time at the same time.  
Another nice thing exhibited by the plot is that the dimension of the observations does not play a significant role in the choice of $k$.   The profiles for different dimensions, from $d=10$ to $d=10000$, are almost the same.  If we increase the strength of the signal (Figure \ref{fig:kchoice50_2}), the whole curve shifts upward, while the profiles for different dimensions still remain the same. 

\begin{figure}[!htp]
\includegraphics[width=0.85\textwidth]{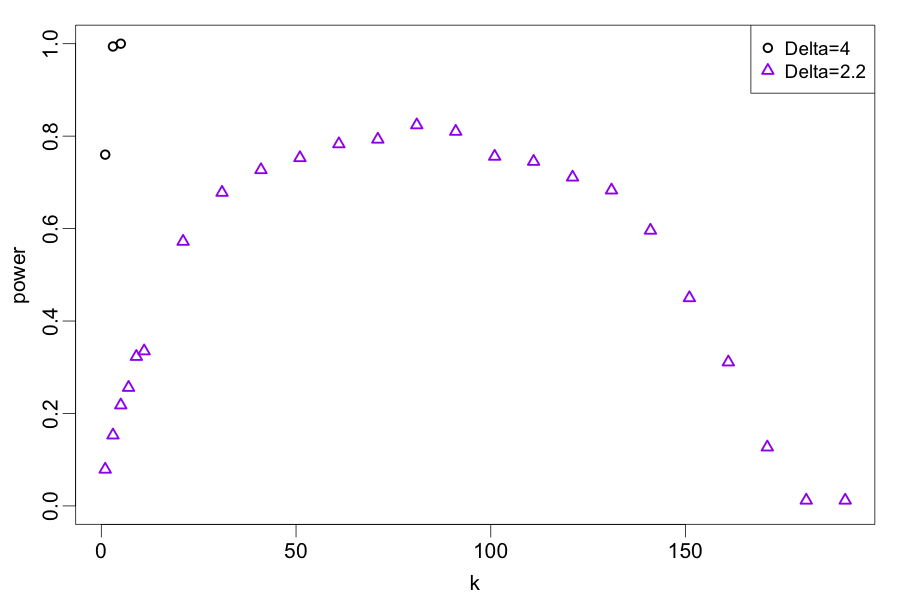}
\caption{The same set up as in Figure \ref{fig:kchoice50} while $L$ is set to be 200.  The dimension of the observations in the sequence is 1000.}\label{fig:kchoice200}
\end{figure}

When we set $L$ to be larger (Figure \ref{fig:kchoice200}, $L=200$, versus Figure \ref{fig:kchoice50}, $L=50$), a similar shape is observed.  It is worthwhile to note that the power of the test increases dramatically as $L$ increase: For $d=1000$, the power achieved by $\Delta=4.5$ for $L=50$ is achieved at $\Delta=2.2$ for $L=200$.  If we set $\Delta=4$ for $L=200$, the power is almost 100\% for 3-NN and 5-NN (shown as circles in Figure \ref{fig:kchoice200}).

In practice, for high-dimensional data or non-Euclidean data, sometimes, only large changes may be of interest, then a relative small $k$ would be preferred as large $k$ may detect small changes.  On the other hand, if all small changes are of interest, then a relatively large $k$ would be recommended.  Also, since the statistics are easy and fast to compute, it might be helpful to run the detection for a number of $k$'s simultaneously.

\subsection{Gradual change}
\label{sec:gradual}

In some applications, the change may happen gradually rather than abruptly.  Even though the proposed method is designed for detecting abrupt changes, it also works for gradual change as long as the change per unit time is relatively strong.  

\begin{figure}[!htp]
\includegraphics[width=.7\textwidth]{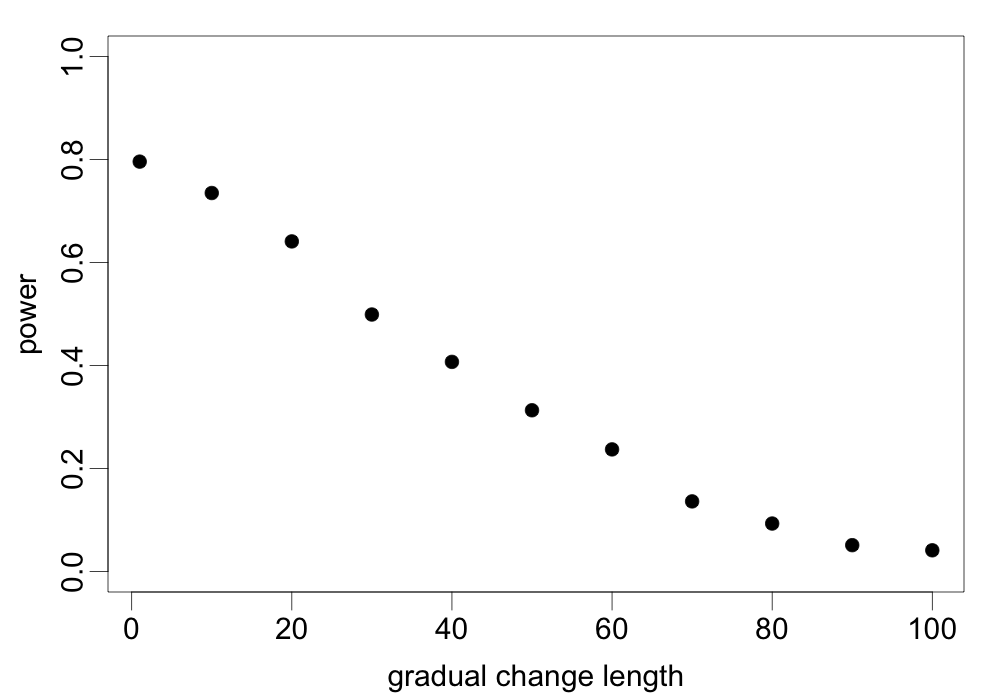}
\caption{The power of the test based on 5-NN for the same amount of change in the mean with the change speed differs.  The `gradual change length' is the amount of time the change takes to finalize.  The longer the gradual change length, the slower the change happens.  The dimension of the observations in the sequence is 1000 and $L$ is set to be 200.  The power is estimated from 1,000 simulation runs and we call the detection successful if it detects the change within 100 observations from the change starts to happen. }\label{fig:gradual}
\end{figure}

Figure \ref{fig:gradual} plots the power of the test based on 5-NN for a change of mean.  In all scenarios, the $L_2$ distance between the mean before the change and the mean after the change is 3.  However, the change could take more than one unit of time to finish.  For example, if the 	`gradual change length' is 10, then $\|\ep(\bY_{\tau+9})-\ep(\bY_{\tau-1})\|_2 = 3$ where $\tau$ is the time the change starts to happen.  For simplicity, we let the change speed to be the same over the gradual change period.  We see that the power decreases as the change takes longer for the same amount of change.    However, the decrease in power is not too bad if the length of the change does not take too long to finalize.  For example, when the 	`gradual  change length' is 20, the power is 0.64, about 80\% of the power if the same amount of change happens abruptly.

\subsection{Possible extensions to other graphs}
In this work, the focus is on the tests based on $k$-NNs.  However, similar tests could be defined for other types of similarity graphs.  For example, we could constructed the minimum spanning tree (MST) constructed on the most recent $L$ observations for each $n$, which is a graph that connects to the most recent $L$ observations with the sum of the distances on the edges minimized, and denote the graph to be $\mathcal{M}_{n_L}$.  Then, $R_L(t,n)$ could be defined as the number of edges in $\mathcal{M}_{n_L}$ connecting observations before $t$ and after $t$, and the standardization could be done correspondingly.   Most of the theoretical treatments in this work could be adopted while we need to figure out the dynamics of the MSTs along time.  In particular, for $\mathcal{M}_{m_L}$ and $\mathcal{M}_{n_L}$, we would need to figure out the expected number of edges that are shared by the two graphs, and the expected number of pairs of edges with one from $\mathcal{M}_{m_L}$ and the other from $\mathcal{M}_{n_L}$ that share a node.  These expectations are not as straightforwardly obtainable as the counterparts in $k$-NN, but they are tractable.  Also, if other ways of the constructing the similarity graph are used rather than the MST, similar arguments follows.   Hence, this current work sets up the basics for graph-based methods for online change-point detection and the special treatments for different similarity graphs are more or less graph-specific.  These specific treatments for other classic similarity graphs will be carried out in future works.

\section{Conclusion}
\label{sec:conclusion}

We propose a new framework for detecting change-points sequentially as data are generated.  Motivated by the complexity of observations in many real applications, we propose to use nearest neighbor information among the observations for sequential detection.  These information can usually be provided by domain experts and thus the proposed method has a wide range of applications.

We explored several stopping rules and the one based on the most recent observations is recommended as it has the desirable property that the detection power is the same across the time.  The asymptotic properties of this stopping rule is studied and the analytic approximation for calculating the average  run length works well for finite samples after skewness correction.  The proposed test exhibits higher power than the parametric method based on normal theory when the dimension of the data is high and/or distributional assumptions for the parametric method are violated.
The proposed method is illustrated on the analysis of friendship network data over time and  some interesting insights are obtained.  


\section*{Acknowledgements}
Hao Chen is support in part by NSF award DMS-1513653.  The author thank David Siegmund, Nancy Zhang, and Jie Peng for helpful discussions.  
%

\bibliographystyle{imsart-nameyear}
\bibliography{online}

\appendix

\section{Proofs for lemmas and theorems}

\subsection{Proof of Theorem \ref{thm:gaussian}}
\label{sec:GaussianProof}
We here prove that $$\{Z_L([vL], [wL]): 0<w-1<v<w<\infty\}$$ converges to a two-dimensional Gaussian random field as $L\rightarrow\infty$.  We only need to show that $$(Z_L([v_1L],[w_1L]), Z_L([v_2L],[w_2L]), \dots, Z_L([v_JL],[w_JL]))$$ converges to a $J$-dimensional Gaussian distribution as $L\rightarrow\infty$ for any $1<w_1<w_2<\dots w_J<\infty$, $v_j\in(w_j-1,w_j),\ j=1,\dots,J$ for any fixed $J$.  
We show in the following for the case $J=2$.  For $J>2$, the proof can be done in the same manner but with a more careful treatment of the indices.  


To begin the proof, we first take a different perspective of $R_L(t,n)$.  Since each permutation can be viewed as another realization of $\bY$, we can re-write $R_L(t,n)$ as
\begin{align*}
R_L(t,n) = 2\sum_{i=n-L+1}^t \sum_{j=t+1}^n (A_{n_L,ij}^+ + A_{n_L,ji}^+).
\end{align*}

For different $i,j,l\in n_L$, we have
\begin{align*}
\pr&(A_{n_L,ij}^+ = 1) = \frac{k}{L-1}, \\
\pr&(A_{n_L,ij}^+ = 1, A_{n_L,ji}^+ = 1) = \frac{p_{k,L}}{L-1}, \\
\pr&(A_{n_L,ji}^+ = 1, A_{n_L,li}^+ = 1) = \frac{q_{k,L}}{(L-1)(L-2)}.
\end{align*}
Since $\sum_{j\in n_L} A_{n_L,ij}^+ = k, \forall i\in n_L$, we have, for different $i,j,l,r\in n_L$,
\begin{align*} 
\pr&(A_{n_L,ij}^+ = 1, A_{n_L,il}^+ = 1) = \frac{k(k-1)}{(L-1)(L-2)}, \\
\pr&(A_{n_L,ij}^+ = 1, A_{n_L,li}^+ = 1) = \frac{k^2 - p_{k,L}}{(L-1)(L-2)}, \\
\pr&(A_{n_L,ij}^+ = 1, A_{n_L,lr}^+ = 1) = \frac{k^2(L-3)+p_{k,L}-q_{k,L}}{(L-1)(L-2)(L-3)}.
\end{align*}
Therefore, $A_{n_L,ij}^+$ and $A_{n_L,lr}^+$ are dependent even when $i,j,l,r$ are all different.  
Hence, all $A_{n_L,ij}^+$'s ($i,j\in n_L,\ i\neq j$) are dependent with each other.  

We consider a set of $\{\widetilde{A}_{n_L,ij}^+\}_{i,j\in n_L}$ that are less dependent than $\{A_{n_L,ij}^+\}_{i,j\in n_L}$. 
For different $i,j,l\in n_L$, let $\{A_{n_L,ij}^+\}_{i,j\in n_L}$ be Bernoulli random variables with
\begin{align*}
\pr&(\widetilde{A}_{n_L,ij}^+=1) = \dfrac{k}{L-1}, \\
\pr&(\widetilde{A}_{n_L,ij}^+ = 1, \widetilde{A}_{n_L,ji}^+ = 1) = \frac{p_{k,L}}{L-1},\\
\pr&(\widetilde{A}_{n_L,ji}^+ = 1, \widetilde{A}_{n_L,li}^+ = 1) = \frac{q_{k,L}}{(L-1)(L-2)}.
\end{align*}
These quantities are defined the same as those for $A_{n_L,ij}^+$'s.
However, we let $\widetilde{A}_{n_L,ij}^+$ be independent of  $\{\widetilde{A}_{n_L,il}^+, \widetilde{A}_{n_L,li}^+\}_{l\neq j}$.  Also, for different $i,j,l,r\in n_L$, we let $\widetilde{A}_{n_L,ij}^+$ and $\widetilde{A}_{L,lr}^+$ be independent.  Then $\{\widetilde{A}_{n_L,ij}^+\}_{i,j\in n_L}$ are only locally dependent.  However, $\sum_j \widetilde{A}_{n_L,ij}^+$'s would no longer be necessarily $k$.  But if we condition on the events $\left\{\sum_j \widetilde{A}_{n_L,ij}^+=k\right\}_{i\in n_L}$, $\{\widetilde{A}_{n_L,ij}^+\}_{i,j\in n_L}$ becomes $\{A_{n_L,ij}^+\}_{i,j\in n_L}$.

We here define 
\begin{align*}
\widetilde{R}_L(t,n) & = 2\sum_{i=n-L+1}^t \sum_{j=t+1}^n (\widetilde{A}_{n_L,ij}^+ + \widetilde{A}_{n_L,ji}^+), \\
\widetilde{Z}_L(t,n) & = \frac{\widetilde{R}_L(t,n)- \ep(\widetilde{R}_L(t,n))}{\sqrt{\Var(\widetilde{R}_L(t,n))}}, \\
\widetilde{N}_L^{(l)}(t,n) & = \sum_{i=n-L+1}^t \sum_{j\in n_L} \widetilde{A}_{n_L,ij}^+, \\
\widetilde{N}_L^{(r)}(t,n) &  = \sum_{i=t+1}^n \sum_{j\in n_L} \widetilde{A}_{n_L,ij}^+, \\
\widetilde{M}_L^{(l)}(t,n) & = \frac{\widetilde{N}_L^{(l)}(t,n) - \ep(\widetilde{N}_L^{(l)}(t,n))}{\sqrt{\Var(\widetilde{N}_L^{(l)}(t,n))}}, \\
\widetilde{M}_L^{(r)}(t,n) & = \frac{\widetilde{N}_L^{(r)}(t,n) - \ep(\widetilde{N}_L^{(r)}(t,n))}{\sqrt{\Var(\widetilde{N}_L^{(r)}(t,n))}}.
\end{align*}
Let $x=L-(n-t)$, we have
\begin{align*}
\ep(\widetilde{R}_L(t,n)) & = \frac{4kx(L-x)}{L-1}, \\
\Var(\widetilde{R}_L(t,n)) & = \frac{4x(L-x)}{L-1}\left(2k+2p_{k,L}+q_{k,L}-\frac{L+2}{L-3}k^2 \right), \\
\ep(\widetilde{N}_L^{(l)}(t,n)) & = kx, \\
\Var(\widetilde{N}_L^{(l)}(t,n)) & = xk-\frac{x^2k^2}{L-1}+\frac{x(x-1)}{L-1}(p_{k,L}+q_{k,L}), \\
\ep(\widetilde{N}_L^{(r)}(t,n)) & = k(L-x), \\
\Var(\widetilde{N}_L^{(r)}(t,n)) & = (L-x)k-\frac{(L-x)^2k^2}{L-1}+\frac{(L-x)(L-x-1)}{L-1}(p_{k,L}+q_{k,L}).
\end{align*}

To show that $Z_L([v_1 L], [w_1 L])$ and $Z_L([v_2 L], [w_2 L])$ are jointly normal, we only need to show the following two lemmas.  For simplicity, let $t_i=[v_iL], n_i=[w_iL], x_i=L-(n_i-t_i), u_i = w_i-v_i$, $i=1,2$.  We thus work under the following condition.
\begin{condition} \label{cond:OL}
$(n_1-t_1), (L-n_1+t_1), (n_2-t_2), (L-n_2+t_2), (n_2-n_1) = O(L).$
\end{condition}

\begin{lemma}\label{lemma:jointnormal}
Under Condition \ref{cond:OL}, we have
\begin{align}\label{eq:ZM}
\left(\widetilde{Z}_L(t_1, n_1), \widetilde{Z}_L(t_2, n_2), \widetilde{M}_L^{(l)}(t_1,n_1), \widetilde{M}_L^{(r)}(t_1,n_1), \widetilde{M}_L^{(l)}(t_2,n_2), \widetilde{M}_L^{(r)}(t_2,n_2)\right)^\prime 
\end{align}
asymptotically jointly normal and the covariance of 
\begin{align}\label{eq:Ms}
\left(\widetilde{M}_L^{(l)}(t_1,n_1), \widetilde{M}_L^{(r)}(t_1,n_1), \widetilde{M}_L^{(l)}(t_2,n_2), \widetilde{M}_L^{(r)}(t_2,n_2)\right)
\end{align}
positive definite.
\end{lemma}

\begin{lemma}\label{lemma:varratio}
Under Condition \ref{cond:OL}, we have
$$\lim_{L\rightarrow\infty} \frac{\Var(\widetilde{R}_L(t_1, n_1))}{\Var(R_L(t_1,n_1))}=c_1, \quad \lim_{L\rightarrow\infty}\frac{\Var(\widetilde{R}_L(t_2, n_2))}{\Var(R_L(t_2,n_2))} = c_2,$$
where $c_1$ and $c_2$ are constants.
\end{lemma}

From Lemma \ref{lemma:jointnormal}, the conditional distribution of $(\widetilde{Z}_L(t_1, n_1), \widetilde{Z}_L(t_2, n_2))^\prime$ given $(\widetilde{M}_L^{(l)}(t_1,n_1), \widetilde{M}_L^{(r)}(t_1,n_1), \widetilde{M}_L^{(l)}(t_2,n_2), \widetilde{M}_L^{(r)}(t_2,n_2))$ is bivariate normal as $L\rightarrow\infty$.  Since the distribution of $(\widetilde{Z}_L(t_1, n_1), \widetilde{Z}_L(t_2, n_2))^\prime$ conditioning on  $(\widetilde{M}_L^{(l)}(t_1,n_1)=0, \widetilde{M}_L^{(r)}(t_1,n_1)=0, \widetilde{M}_L^{(l)}(t_2,n_2)=0, \widetilde{M}_L^{(r)}(t_2,n_2)=0)$ is equivalent to that conditioning on $\sum_{j\in n_{1,L}}A_{L,ij}^+(n_1) = k, \forall i\in n_{1,L}$ and $\sum_{j\in n_{2,L}}A_{L,ij}^+(n_2) = k, \forall i\in n_{2,L}$.  Together with Lemma \ref{lemma:varratio} and $$\ep(\widetilde{R}_L(t_1, n_1)) = \ep(R_L(t_1,n_1)),\ \ep(\widetilde{R}_L(t_2, n_2)) = \ep(R_L(t_2,n_2)),$$ we have that $(Z_L(t_1,n_1), Z_L(t_2,n_2))^\prime$ becomes bivariate normal as $L\rightarrow\infty$. 

 It is easy to show Lemma \ref{lemma:varratio} as
\begin{align*}
\lim_{L\rightarrow\infty} \frac{\Var(\widetilde{R}_L(t_1, n_1))}{\Var(R_L(t_1,n_1))} = \frac{2k+2q_{1,\infty}+q_{2,\infty}-k^2}{k+4u_1(1-u_1)q_{1,\infty}+(1-2u_1)^2(q_{2,\infty}-k^2)}, \\
\lim_{L\rightarrow\infty} \frac{\Var(\widetilde{R}_L(t_1, n_1))}{\Var(R_L(t_1,n_1))} = \frac{2k+2q_{1,\infty}+q_{2,\infty}-k^2}{k+4u_2(1-u_2)q_{1,\infty}+(1-2u_2)^2(q_{2,\infty}-k^2)}.
\end{align*}
The two limits are constants.

 In the following, we prove Lemma \ref{lemma:jointnormal}.  It is easy to see that the covariance of \eqref{eq:Ms} is positive definite under Condition \ref{cond:OL}.
 To show that \eqref{eq:ZM} is jointly normal, we only need to show that $a_1\widetilde{Z}_L(t_1, n_1)+a_2 \widetilde{Z}_L(t_2, n_2)+a_3 \widetilde{M}_L^{(l)}(t_1,n_1)+a_4 \widetilde{M}_L^{(r)}(t_1,n_1)+a_5 \widetilde{M}_L^{(l)}(t_2,n_2)+a_6 \widetilde{M}_L^{(r)}(t_2,n_2)$ is normal for any fixed $a_1,\dots, a_6$.  Let $\sigma_0^2$ be the variance of this summation.  If $\sigma_0=0$, the quantity is degenerating.  We show in the following the case when $\sigma_0>0$ that the quantity 
\begin{align*}
 W&=\frac{1}{\sigma_0}\left(a_1\widetilde{Z}_L(t_1, n_1)+a_2 \widetilde{Z}_L(t_2, n_2)+a_3 \widetilde{M}_L^{(l)}(t_1,n_1)\right. \\
 & \quad \quad \quad \quad \quad \left. +a_4 \widetilde{M}_L^{(r)}(t_1,n_1)+a_5 \widetilde{M}_L^{(l)}(t_2,n_2)+a_6 \widetilde{M}_L^{(r)}(t_2,n_2)\right)
\end{align*}
 is normal through Stein's method.

 Consider sums of the form
$W=\sum_{i\in{\mathcal{I}}} \xi_i,$
where $\mathcal{I}$ is an index set and $\xi$ are random variables with $\ep(\xi_i)=0$, and $\ep(W^2)=1$.  The following assumption restricts the dependence between $\{\xi_i:~i \in \mathcal{I}\}$.
\begin{assumption} \cite[p.\, 17]{chen2005stein}
  \label{assump:LD}
For each $i\in{\mathcal{I}}$ there exists $S_i \subset T_i \subset {\mathcal{I}}$ such that $\xi_i$ is independent of $\xi_{S_i^c}$ and $\xi_{S_i}$ is independent of $\xi_{T_i^c}$.
\end{assumption}
We will use the following specific form of Stein's method.
\begin{theorem}\label{thm:3.4} \cite[Theorem 3.4]{chen2005stein}
Under Assumption \ref{assump:LD}, we have
$$\sup_{h\in Lip(1)} |\ep h(W) - \ep h(Z)| \leq \delta$$
where $Lip(1) = \{h: \mathbb{R}\rightarrow \mathbb{R}, \|h'\|\leq 1 \}$, $Z$ has ${\cal N}(0,1)$ distribution and
 $$\delta = 2 \sum_{i\in{\mathcal{I}}} (\ep|\xi_i \eta_i\theta_i| + |\ep(\xi_i\eta_i)| \ep|\theta_i|) + \sum_{i\in{\mathcal{I}}} \ep|\xi_i\eta_i^2|$$
with $\eta_i = \sum_{j\in S_i}\xi_j$ and $\theta_i = \sum_{j\in T_i} \xi_j$, where $S_i$ and $T_i$ are defined in Assumption \ref{assump:LD}.
\end{theorem}

Let 
$$\sigma_1 = \sqrt{\Var(\widetilde{R}_L(t_1, n_1))}, \sigma_1^{(l)}=\sqrt{\Var(\widetilde{N}_L^{(l)}(t_1,n_1))}, \sigma_1^{(r)}=\sqrt{\Var(\widetilde{N}_L^{(r)}(t_1,n_1))},$$ $$\sigma_2 = \sqrt{\Var(\widetilde{R}_L(t_2, n_2))}, \sigma_2^{(l)}=\sqrt{\Var(\widetilde{N}_L^{(l)}(t_2,n_2))}, \sigma_2^{(r)}=\sqrt{\Var(\widetilde{N}_L^{(r)}(t_2,n_2))}.$$ 
Then $\sigma_1, \sigma_1^{(l)}, \sigma_1^{(r)}, \sigma_2, \sigma_2^{(l)}, \sigma_2^{(r)} \sim \mathcal{O}(L^{1/2})$

In the following, we write $A_{n_L,ij}^+$ as $A_{L,ij}^+(n)$ to avoid the cumbersome of the subscripts.
Let 
\begin{align*}
\xi_{ij,1} & = \left\{\begin{array}{ll}  \frac{a_3}{\sigma_0\sigma_1^{(l)}}  \left(\widetilde{A}_{L,ij}^+(n_1) -\frac{k}{L-1}\right) & i,j\in\{n_1-L+1,\dots, t_1\}, i\neq j,  \\  \frac{2a_1\sigma_1^{(l)} + a_3 \sigma_1}{\sigma_0\sigma_1\sigma_1^{(l)}}\left(\widetilde{A}_{L,ij}^+(n_1) -\frac{k}{L-1}\right) & i\in\{n_1-L+1,\dots, t_1\}, \\
& j\in\{t_1+1,\dots,n_1\},\\
\frac{2a_1\sigma_1^{(r)}+a_4\sigma_1}{\sigma_0\sigma_1\sigma_1^{(r)}} \left(\widetilde{A}_{L,ij}^+(n_1) -\frac{k}{L-1}\right) & i\in\{t_1+1,\dots,n_1\},\\
& j\in\{n_1-L+1,\dots, t_1\},\\
\frac{a_4}{\sigma_0\sigma_1^{(r)}} \left(\widetilde{A}_{L,ij}^+(n_1) -\frac{k}{L-1}\right), & i, j\in\{n_1-L+1,\dots, t_1\} i\neq j, \\
0 & \text{otherwise},
 \end{array} \right. \\
\xi_{ij,2} & = \left\{\begin{array}{ll}  \frac{a_5}{\sigma_0\sigma_2^{(l)}}  \left(\widetilde{A}_{L,ij}^+(n_2) -\frac{k}{L-1}\right) & i,j\in\{n_2-L+1,\dots, t_2\}, i\neq j,  \\ \frac{2a_2\sigma_2^{(l)}+a_5\sigma_2}{\sigma_0\sigma_2\sigma_2^{(l)}} \left(\widetilde{A}_{L,ij}^+(n_2) -\frac{k}{L-1}\right) & i\in\{n_2-L+1,\dots, t_2\},\\
& j\in\{t_2+1,\dots,n_2\},\\ \frac{2a_2\sigma_2^{(r)}+a_6\sigma_2}{\sigma_0\sigma_2\sigma_2^{(r)}} \left(\widetilde{A}_{L,ij}^+(n_2) -\frac{k}{L-1}\right)  & i\in\{t_2+1,\dots,n_2\},\\
& j\in\{n_2-L+1,\dots, t_2\},\\
\frac{a_6}{\sigma_0\sigma_2^{(r)}} \left(\widetilde{A}_{L,ij}^+(n_2) -\frac{k}{L-1}\right) & i, j\in\{n_2-L+1,\dots, t_2\}, i\neq j, \\
0 & \text{otherwise},
 \end{array} \right. \\
\xi_{ij} & = \xi_{ij,1} + \xi_{ij,2}.
\end{align*}
Then
$ W = \sum_{i,j} \xi_{ij}$ and $\xi_{ij}$ is only dependent with $\{\xi_{ji},\xi_{lj},\forall l\}$.
Adopting the same notations in Theorem \ref{thm:3.4}, we have
\begin{align*}
\eta_{ij} & = \xi_{ij} + \xi_{ji} + \sum_{l} \xi_{lj}, \\
\theta_{ij} & = \xi_{ij} + \xi_{ji} + \sum_{l} (\xi_{lj} + \xi_{jl} + \xi_{li}).
\end{align*}
Let 
\begin{align*}
a & =\max(|a_3|, |2a_1+a_3|, |2a_1+a_4|, |a_4|, |a_5|,|2a_2+a_5|,|2a_2+a_6|,|a_6|), \\
\sigma & = \sigma_0 \times \min(\sigma_1,\sigma_1^{(l)},\sigma_1^{(r)},\sigma_2,\sigma_2^{(l)},\sigma_2^{(r)}),
\end{align*} then $a=O(1), \sigma=O(L^{1/2})$, and $|\xi_{ij}|\leq \dfrac{2ak}{\sigma}$.

Let $\widetilde{N}_1(i) = \sum_{j\in n_{1,L}} \widetilde{A}_{L,ij}^+(n_1)$,  $\widetilde{N}_2(i) = \sum_{j\in n_{2,L}} \widetilde{A}_{L,ij}^+(n_2)$.  Then, under Condition \ref{condition:degree}, we have
\begin{align*}
-k\frac{a}{\sigma} \leq \sum_l \xi_{jl,1} \leq \widetilde{N}_1(j)\frac{a}{\sigma}, \\
-k\frac{a}{\sigma} \leq \sum_l \xi_{lj,1} \leq \mathbb{C}\frac{a}{\sigma}, \\
-k\frac{a}{\sigma} \leq \sum_l \xi_{jl,2} \leq \widetilde{N}_2(j)\frac{a}{\sigma}, \\
-k\frac{a}{\sigma} \leq \sum_l \xi_{lj,2} \leq \mathbb{C}\frac{a}{\sigma}.
\end{align*}
So
\begin{align*}
\sum_{l} (\xi_{lj} + \xi_{jl} + \xi_{li}) & = \sum_{l} (\xi_{lj,1} + \xi_{jl,1} + \xi_{li,1} + \xi_{lj,2} + \xi_{jl,2} + \xi_{li,2}) \\
& \in\left[-6k, \widetilde{N}_1(j)+\widetilde{N}_2(j) + 4\mathbb{C}  \right] \times\frac{a}{\sigma}.
\end{align*}
Hence, $|\theta_{ij}| \leq (\widetilde{N}_1(j)+\widetilde{N}_2(j) + 4\mathbb{C}  + 10k) \dfrac{a}{\sigma}$.  Similarly $|\eta_{ij}| \leq (2\mathbb{C}  + 6k) \dfrac{a}{\sigma}$. 

Since 
\begin{align*}
& \left|\ep\left(\left(A_{L,ij}^+(n_r)-\frac{k}{L-1}\right)\left(A_{L,i^\prime j^\prime}^+(n_{r^\prime})-\frac{k}{L-1}\right)\right) \right| \leq \frac{2k}{L-1},  \\
& \quad \quad \quad  \quad \quad \quad \quad \quad  \quad \quad \quad \quad \quad  \quad \quad \forall i,j,i^\prime, j^\prime, \text{ and } r, r^\prime\in\{1,2\},
\end{align*}
we have $|\ep(\xi_{ij}\xi_{lr})| \leq \dfrac{4a^2}{\sigma^2}\dfrac{2k}{L-1}$, so
\begin{align*}
& |\ep(\xi_{ij}\eta_{ij})| \\
& \quad = |\ep(\xi_{ij}(\xi_{ij} + \xi_{ji} + \sum_{l} \xi_{lj}))|  \leq \dfrac{4a^2}{\sigma^2}\dfrac{2k}{L-1} (2+\widetilde{N}_1(j)+\widetilde{N}_2(j)), \\
& |\ep(\xi_{ij}\eta_{ij}\theta_{ij})| \\
& \quad \leq |\ep((\widetilde{N}_1(j)+\widetilde{N}_2(j) + 4\mathbb{C}  + 10k) \dfrac{a}{\sigma}\dfrac{4a^2}{\sigma^2}\dfrac{2k}{L-1} (2+\widetilde{N}_1(j)+\widetilde{N}_2(j))) |\\
& \quad = \frac{8a^3 k}{\sigma^3 (L-1)}(4\mathbb{C}+12k+\ep((\widetilde{N}_1(j)+\widetilde{N}_2(j))^2) + 2k(4\mathbb{C}+10k)) \\
& \quad \leq  \frac{8a^3 k}{\sigma^3 (L-1)}(4\mathbb{C}+12k+4(k^2+k) + 2k(4\mathbb{C}+10k)), \\
& |\ep(\xi_{ij}\eta_{ij}^2)| \\
& \quad \leq |\ep((2\mathbb{C}  + 6k) \dfrac{a}{\sigma}\dfrac{4a^2}{\sigma^2}\dfrac{2k}{L-1} (2+\widetilde{N}_1(j)+\widetilde{N}_2(j))) |\\
& \quad = \frac{8a^3 k}{\sigma^3 (L-1)}(2\mathbb{C}+6k)(2+2k).
\end{align*}
Hence
\begin{align*}
\delta & \leq  \frac{40a^3 k}{\sigma^3 (L-1)}(28k^2 + 16\mathbb{C}+8k\mathbb{C}) L^2 =O(L^{-1/2})\overset{L\rightarrow\infty}{\rightarrow} 0.
\end{align*}

\subsection{Proof of Theorem \ref{thm:h12}}
\label{sec:proofthmh12}

Let $\rho_{L,(s,m)}(t,n) = \Cov(Z_L(s,m), Z_L(t,n))$, then 
\begin{align*}
& \rho^\star_{((m-s)/L, m/L)}((t-s)/L, (n-m)/L) = \lim_{L\rightarrow\infty} \rho_{L,(s,m)}(t,n), \\
& \left. \frac{\partial_+ \rho^\star_{(u,w)}(\delta_1,0)}{\partial \delta_1}\right|_{\delta_1=0}  =  \lim_{L\rightarrow\infty} L \left.\frac{\partial \rho_{(s,m)}(t,m)}{\partial t}\right|_{t\searrow s}, \\
& \left. \frac{\partial_- \rho^\star_{(u,w)}(\delta_1,0)}{\partial \delta_1}\right|_{\delta_1=0}  =  \lim_{L\rightarrow\infty} L \left.\frac{\partial \rho_{(s,m)}(t,m)}{\partial t}\right|_{t\nearrow s}, 
\end{align*}
and similar relations for the two directional partial derivatives with respect to $\delta_2$. 

Without loss of generality, let $m\leq n$. 
Since we are only interested in these directional partial derivatives, we only need to figure out the covariance for two scenarios: (i) $s<t<m\leq n$, and (ii) $t<s<m\leq n$, under $(n-m)=o(L)$.  

We have 
\begin{align*}
\Cov(Z_L(s,m), Z_L(t,n))=\frac{\ep(R_L(s,m)R_L(t,n))-\ep(R_L(s,m))\ep(R_L(t,n))}{\sqrt{\Var(R_L(s,m))\Var(R_L(t,n))}}.
\end{align*}
We know that 
\small
\begin{align}
& \ep(R_L(t,n)) = \frac{4k(n-t)(L-n+t)}{L-1}, \label{eq:ERL} \\
& \Var(R_L(t,n)) = \frac{4(n-t)(L-n+t)}{L-1}  \label{eq:VRL} \\
& \quad \times \left(\frac{4(n-t-1)(L-n+t-1)}{(L-2)(L-3)}  \left(p_{k,L} -q_{k,L} + \frac{(L-3)k^2}{L-1} \right) + \left(q_{k,L}+k - k^2\right)  \right), \nonumber
\end{align}
\normalsize
and similar equations for $\ep(R_L(s,m)$ and $\Var(R_L(s,m))$. 
So what we need to calculate is $\ep(R_L(s,m) R_L(t,n))$ under the two scenarios.

\noindent \textbf{(i) $s<t<m\leq n$}

Notice that we can rewrite $R_L(s,m)$ and $R_L(t,n)$ to be
\begin{align*}
R_L(s,m) & = \sum_{i,j\in m_L} (A_{m_L,ij}^+ + A_{m_L,ji}^+)B_{ij}(s,m_L\cup n_L), \\
R_L(t,n) & = \sum_{i,j\in n_L} (A_{n_L,ij}^+ + A_{n_L,ji}^+ )B_{ij}(t,m_L\cup n_L).
\end{align*}
For $i,j\in m_L\cap n_L$, we have
\begin{align} \label{eq:EB1}
\ep(B_{ij}(s,m_L\cup n_L)B_{ij}(t,m_L\cup n_L))  = \frac{2(L-(n-s))(m-t)}{(L-(n-m))(L-(n-m)-1)}.
\end{align}
For $i\in m_L\cap n_L,\ j\in m_L,\ l\in n_L$, and $l\neq j$, we have
\begin{align} \label{eq:EB2}
&\ep(B_{ij}(s,m_L\cup n_L)B_{il}(t,m_L\cup n_L))  \\
& = \frac{(L-(n-s))[(m-t)(L-2)+(n-m)(m-s)]}{(L-(n-m))[(L-(n-m)-1)(L-2)+(n-m)(L-1)]}  \nonumber \\
& \quad + \frac{(t-s)(L-(m-s))(n+m-2t)}{(L-(n-m))[(L-(n-m)-1)(L-2)+(n-m)(L-1)]}. \nonumber
\end{align}
For $i,j\in m_L,\ l,r\in n_L,\ l\neq i,j$ and $r\neq i,j$, we have
\begin{align} \label{eq:EB3}
& \ep(B_{ij}(s,m_L\cup n_L)B_{lr}(t,m_L\cup n_L)) \\
& = \frac{4(L-(m-s))(L-(n-t)-1)[(n-t)(m-s)-(m-t)]}{L(L-1)(L-2)(L-3)+2(n-m)(2L^2-6L+3)+2(n-m)^2} \nonumber \\
& \quad +\frac{4[(n-m)(m-t)(n-t-1) - (L-(n-s))(t-s)(n-t)]}{L(L-1)(L-2)(L-3)+2(n-m)(2L^2-6L+3)+2(n-m)^2}. \nonumber
\end{align}
We denote the quantities in \eqref{eq:EB1}, \eqref{eq:EB2} and \eqref{eq:EB3} by $f_1(s,t), f_2(s,t)$ and $f_3(s,t)$, respectively.
Then
\begin{align*}
\ep &(R_L(s,m)R_L(t,n)) = \ep(\ep(R_L(s,m)R_L(t,n)|\bY)) \\
 & = f_1(s,t)\  \ep\left( \sum_{i,j\in m_L\cap n_L} (A_{m_L,ij}^+ + A_{m_L,ji}^+)(A_{n_L,ij}^+ + A_{n_L,ji}^+)\right)  \\
& \quad + f_2(s,t) \ \ep\left( \sum_{i\in m_L\cap n_L; j\in m_L, l\in n_L; l\neq j} (A_{m_L,ij}^+ + A_{m_L,ji}^+)(A_{n_L,il}^+ + A_{n_L,li}^+)\right)   \\
& \quad + f_3(s,t)\ \ep\left( \sum_{i,j\in m_L; l,r\in n_L; l\neq i,j; r\neq i,j} (A_{m_L,ij}^+ + A_{m_L,ji}^+)(A_{n_L,lr}^+ + A_{n_L,rl}^+)\right) .
\end{align*}

Let
\begin{align}
X_1 & = \sum_{i,j\in m_L\cap n_L} (A_{m_L,ij}^+ + A_{m_L,ji}^+)(A_{n_L,ij}^+ + A_{n_L,ji}^+) \label{eq:X1} \\
& = 2\sum_{i,j\in m_L\cap n_L}  (A_{m_L,ij}^+ A_{n_L,ij}^+ + A_{m_L,ij}^+ A_{n_L,ji}^+), \nonumber\\
X_2 & = \sum_{i\in m_L\cap n_L; j\in m_L, l\in n_L} (A_{m_L,ij}^+ + A_{m_L,ji}^+)(A_{n_L,il}^+ + A_{n_L,li}^+), \label{eq:X2}\\
& = (L-(n-m))k^2 + k\sum_{i\in m_L\cap n_L} D_{n_L}(i) + k\sum_{i\in m_L\cap n_L} D_{m_L}(i) \nonumber  \\
& \quad \quad + \sum_{i\in m_L\cap n_L} D_{m_L}(i)D_{n_L}(i), \nonumber \\
X_3 & = \sum_{i,j\in m_L; l,r\in n_L} (A_{m_L,ij}^+ + A_{m_L,ji}^+)(A_{n_L,lr}^+ + A_{n_L,rl}^+) = 4L^2k^2, \nonumber 
\end{align}
where $D_{m_L}(i) = \sum_{j} A_{m_L,ji}^+$, $D_{n_L}(i) = \sum_{j} A_{n_L,ji}^+$.

Then
\begin{align*}
& \sum_{i\in m_L\cap n_L; j\in m_L, l\in n_L; l\neq j} (A_{m_L,ij}^+ + A_{m_L,ji}^+)(A_{n_L,il}^+ + A_{n_L,li}^+)  = X_2-X_1, \\
& \sum_{i,j\in m_L; l,r\in n_L; l\neq i,j; r\neq i,j}  (A_{m_L,ij}^+ + A_{m_L,ji}^+)(A_{n_L,lr}^+ + A_{n_L,rl}^+) \\
& \quad \quad \quad \quad \quad \quad \quad \quad \quad  = X_3-2X_1-4(X_2-X_1) = X_3+2X_1-4X_2,
\end{align*}
%
%
and 
\begin{align}\label{eq:Est}
\ep(R_L(s,m)R_L(t,n)&) = (2f_1(s,t)-4f_2(s,t)+2f_3(s,t))\ep(X_1) \\
&\ + (4f_2(s,t)-4f_3(s,t))\ep(X_2) + 4L^2k^2f_3(s,t). \nonumber
\end{align}

\noindent \textbf{(ii) $t<s<m\leq n$}

Following the same argument, we have 
\begin{align}\label{eq:Ets}
\ep(R_L(s,m)R_L(t,n)) & = (2f_4(s,t)-4f_5(s,t)+2f_6(s,t))\ep(X_1) \\
&+ (4f_5(s,t)-4f_6(s,t))\ep(X_2) + 4L^2k^2f_6(s,t), \nonumber
\end{align}
where
\begin{align*}
f_4(s,t) & = \frac{2(L-(n-t))(m-s)}{(L-(n-m))(L-(n-m)-1)}, \\
f_5(s,t) & = \frac{(L-(n-t))(m-s)(L+n-m+2s-2t-2)}{(L-(n-m))((L-(n-m)-1)(L-2)+(n-m)(L-1))}, \\
f_6(s,t) 
& = \frac{4(L-(n-t))(m-s)[(n-t-1)(L-(m-s)-1)-(s-t)]}{L(L-1)(L-2)(L-3)+2(n-m)(2L^2-6L+3)+2(n-m)^2}.
\end{align*}

So for both scenarios, the problem boils down to calculate $\ep(X_1)$ and $\ep(X_2)$. 
It is easy to have that 
\begin{align}\label{eq:EDi}
\ep\left(\sum_{i\in m_L\cap n_L} D_{n_L}(i)\right) & = \ep\left(\sum_{i\in m_L\cap n_L} D_{m_L}(i)\right) = k(L-(n-m)).
\end{align}
In the following, we calculate the remaining three quantities:  
\begin{align}
\label{eq:Eq0mn} \ep\left(\sum_{i,j\in m_L\cap n_L} A_{m_L,ij}^+  A_{n_L,ij}^+\right), \\
\label{eq:Eq1mn} \ep\left(\sum_{i,j\in m_L\cap n_L} A_{m_L,ij}^+  A_{n_L,ji}^+\right), \\
\label{eq:Eq2mn} \ep\left(\sum_{i\in m_L\cap n_L} D_{m_L}(i) D_{n_L}(i)\right).
\end{align}
We calculate them under $(n-m)=o(L)$.


For \eqref{eq:Eq0mn}, we have
\begin{align*}
& \ep\left(\sum_{i,j\in m_L\cap n_L} A_{m_L,ij}^+ A_{n_L,ij}^+\right) \\
& \quad = (L-(n-m))(L-(n-m)-1) \pr (A_{m_L,ij}^+ = 1, A_{n_L,ij}^+ = 1).
\end{align*}
For $i,j\in m_L\cap n_L,\ i\neq j$, when $k=1$, $$\pr (A_{m_L,ij}^+ = 1, A_{n_L,ij}^+ = 1)=\pr (A_{m_L\cup n_L,ij}^+ = 1) = \frac{1}{L+n-m-1}.$$
When $k>1$,
\begin{align*}
\pr&(A_{m_L,ij}^+ = 1, A_{n_L,ij}^+ = 1) \\
& = \pr(A_{m_L\cup n_L,ij}^+ = 1) + \pr(A_{m_L,ij}^+ = 1, A_{n_L,ij}^+ = 1, A_{m_L\cup n_L,ij}^+ = 0) \\
& = \frac{k}{L+n-m-1} + O\left(\frac{(n-m)^2}{L^3}\right).
\end{align*}
The event $\{A_{m_L,ij}^+ = 1, A_{n_L,ij}^+ = 1, A_{m_L\cup n_L,ij}^+ = 0\}$ implies that there is at least one observation in $\{m-L+1,\dots,n-L\}$ and another observation in $\{m+1,\dots,n\}$ whose distance to $\bY_i$ is smaller than the distance between $\bY_i$ and $\bY_j$. The order of its probability follows by considering the possible locations of the first $k$ NNs of $Y_i$ among all observations in $m_L\cup n_L$.
So
\begin{align}\label{eq:AijAij}
& \ep\left(\sum_{i,j\in m_L\cap n_L} A_{m_L,ij}^+ A_{n_L,ij}^+\right) \\
& = \frac{k(L-(n-m))(L-(n-m)-1)}{L+n-m-1} + (k-1)~O\left( \frac{(n-m)^2}{L} \right) \nonumber \\
& = k\left(L-3(n-m) \right) + O\left( \frac{(n-m)^2}{L} \right). \nonumber
\end{align}

For \eqref{eq:Eq1mn}, notice that, for $i,j\in m_L\cap n_L,\ i\neq j$,
\begin{align*}
\pr&(A_{m_L,ij}^+ = 1, A_{n_L,ji}^+ = 1) \\
& = \pr(A_{m_L\cup n_L,ij}^+ =1, A_{m_L\cup n_L,ji}^+ = 1) \\
& \quad + \pr(A_{n_L,ji}^+ = 1, A_{m_L\cup n_L,ij}^+ =1, A_{m_L\cup n_L,ji}^+ = 0) \\
& \quad + \pr(A_{m_L,ij}^+ = 1, A_{m_L\cup n_L,ij}^+ =0, A_{m_L\cup n_L,ji}^+ = 1) \\
& \quad + \pr(A_{m_L,ij}^+ = 1, A_{n_L,ji}^+ = 1, A_{m_L\cup n_L,ij}^+ =0, A_{m_L\cup n_L,ji}^+ = 0) \\
& = \frac{2 p_{k,L}}{L-1} - \frac{q_{1,L+n-m}^{(k)}}{L+n-m-1} - \frac{2(n-m)\sum_{r=1}^k p_{k,L}(k,r))}{(L-1)(L+n-m-2)} +  O\left(\frac{(n-m)^2}{L^3}\right).
\end{align*}
The last step can be obtained by considering all possible locations of the first $k$ NNs of $\bY_i$ among all observations in $m_L\cup n_L$, and all possible locations of the first $k$ NNs of $\bY_j$ among all observations in $m_L\cup n_L$.
Then, we have, after simplification,
\begin{align}\label{eq:AijAji}
 & \ep\left(\sum_{i,j\in m_L\cap n_L} A_{m_L,ij}^+ A_{n_L,ji}^+\right) 
  = L\left(2 p_{k,L}-p_{k,L+n-m}\right)\\
  -&(n-m)\left(4p_{k,L}-3p_{k,L+n-m}+2\sum_{r=1}^k p_{k,L}(k,r)\right) + O\left( \frac{(n-m)^2}{L} \right). \nonumber
\end{align}

For \eqref{eq:Eq2mn}, we have
\begin{align*}
& \ep\left(\sum_{i\in m_L\cap n_L} D_{m_L}(i) D_{n_L}(i)\right) = \ep\left(\sum_{i\in m_L\cap n_L; j\in m_L; l\in n_L} A_{m_L,ji}^+A_{n_L,li}^+\right) \\ 
& = \ep\left(\sum_{i,j\in m_L\cap n_L}A_{m_L,ji}^+A_{n_L,ji}^+ \right)+ \ep\left(\sum_{i\in m_L\cap n_L; j\in m_L; l\in n_L; l\neq j} A_{m_L,ji}^+A_{n_L,li}^+\right).
\end{align*}
The first part of the summation is \eqref{eq:Eq0mn}.  For the second part, notice that, for $i\in m_L\cap n_L,\ j\in m_L,\ l\in n_L,\ l\neq j$,
\begin{align*}
& \pr(A_{m_L,ji}^+ = 1, A_{n_L,li}^+ = 1)  \\
& = \pr(A_{m_L\cup n_L,ji}^+ = 1, A_{m_L\cup n_L,li}^+ = 1) \\
& \quad + \pr(A_{m_L,ji}^+ = 1, A_{m_L\cup n_L,ji}^+ = 0, A_{m_L\cup n_L,li}^+ = 1)\\
& \quad + \pr(A_{n_L,li}^+ = 1, A_{m_L\cup n_L,ji}^+ = 1, A_{m_L\cup n_L,li}^+ = 0) \\
& \quad + \pr(A_{m_L,ji}^+ = 1, A_{n_L,li}^+ = 1, A_{m_L\cup n_L,ji}^+ = 0, A_{m_L\cup n_L,li}^+ = 0) \\
& = \frac{2 q_{k,L}}{(L-1)(L-2)} - \frac{q_{k,L+n-m}}{(L+n-m-1)(L+n-m-2)}  \\
& \quad \quad - \frac{2(n-m)\sum_{r=1}^k q_{L}(k,r))}{(L-1)(L-2)(L+n-m-2)} +  O\left(\frac{(n-m)^2}{L^4}\right).
\end{align*}
The last step can be obtained by considering all possible locations of the first $k$ NNs of $\bY_j$ among all observations in $m_L\cup n_L$, and all possible locations of the first $k$ NNs of $\bY_l$ among all observations in $m_L\cup n_L$.
Then, we have, after simplification,
\begin{align}\label{eq:AjiAli}
& \ep\left(\sum_{i\in m_L\cap n_L} D_{m_L}(i) D_{n_L}(i)\right)
= L\left(k+2q_{k,L} - q_{k,L+n-m} \right)  \\
  +(n-&m)\left( 3q_{k,L+n-m}-2q_{k,L}-3k  -2\sum_{r=1}^k q_{L}(k,r)) \right)  + O\left(\frac{(n-m)^2}{L}\right). \nonumber
\end{align}

Plugging \eqref{eq:EDi}, \eqref{eq:AijAij}, \eqref{eq:AijAji} and \eqref{eq:AjiAli} into \eqref{eq:Est} and \eqref{eq:Ets}, we get an analytic expression of the covariance between $Z_L(s,m)$ and $Z_L(t,n)$.  Taking partial derivatives over $t$ and $n$, respectively, and then taking the limit, after some tedious calculations, we get Theorem \ref{thm:h12}.


\end{document}